\documentclass{JHEP}
\input epsf
\usepackage{epsfig}
\usepackage{amssymb}
\usepackage[active]{srcltx}
\usepackage{comment}
\usepackage{bm}
\newcommand{\bea}{\begin{eqnarray}}
\newcommand{\eea}{\end{eqnarray}}
\newcommand{\bean}{\begin{eqnarray*}}
\newcommand{\eean}{\end{eqnarray*}}
\newcommand{\nn}{\nonumber \\}

\def\W #1{\widetilde{#1}}

\def\braket#1{\left\langle #1 \right\rangle}

\def\ket#1{\left| #1\right\rangle}
\def\gb #1{ \left\langle #1 \right]}

\def\det{\mathop{\rm det}}

\def\eref#1{(\ref{#1})}

\def\wt{\widetilde}

\def\a{{\alpha}}

\def\b{{\beta}}

\def\eps{\epsilon}

\def\vev{\braket}

\def\bket#1{\left| #1\right]}
\def\bvev#1{\left[ #1 \right]}
\def\Spaa{\vev}
\def\Spbb{\bvev}
\def\Spab{\gb}

\def\Label#1{\label{#1}%
  \smash{\hbox to0pt{\raise1ex\hbox{\tiny[#1]}\hss}}}

\preprint{
\\
{\tt hep-th/}}
\title{Analytic structure of one-loop coefficients}
\author{Bo Feng$^{a,b}$, Honghui Wang$^{a}$\\$^a$Zhejiang Institute of Modern Physics, Zhejiang
University, Hangzhou, 310027, P. R. China\\$^b$Center of
Mathematical Sciences, Zhejiang University, Hangzhou, 310027, P. R.
China}
\abstract{By the unitarity cut method, analytic expressions of
one-loop coefficients have been given in spinor forms. In this
paper, we present one-loop coefficients of various bases in
Lorentz-invariant contraction forms of external momenta. Using these
forms, the analytic structure of these coefficients becomes
manifest. Firstly,  coefficients of bases contain only second-type
singularities while the first-type singularities are included inside scalar
bases. Secondly, the highest degree of each singularity is correlated
with the degree of the inner momentum in the numerator. Thirdly,
the same singularities will appear in different coefficients, thus
our explicit results could be used to provide a clear
physical  picture under various limits (such as soft or
collinear limits) when combining contributions from all bases.
}
\keywords{Analytic structure, one-loop coefficients}

\begin{document}

\section{Introduction}

In  the last ten years, enormous progresses have been made in the
computation of scattering amplitudes at both tree-level (see
\cite{Feng:2011np, Brandhuber:2011ke})  and one-loop level (see, for
example, the reference \cite{Bern:2008ef, AlcarazMaestre:2012vp,
Binoth:2010ra} and citations in the paper). All these progresses
come out as a better understanding of the analytic structure of
scattering amplitudes at various orders.

At the tree-level, the analytic structure is relatively simple:
there is only the single-pole structure. However, even with the single-pole
structure, since there are many kinematic variables, we are facing
the multi-variable complex analysis  as shown in the old S-matrix
program \cite{S-matrix}, whose central theme is to determining
scattering amplitudes directly from their analytic structures. This
complicated mathematical problem is avoided in the BCFW on-shell
recursion relations
\cite{Britto:2004ap,Britto:2005fq,Bedford:2005sf,Cachazo:2005sf,
Benincasa:2007ff,Nima:2008sj,Cheung:2008fa}, which
is an outgrowth of Witten's twistor
program\cite{Witten:2003nn,Cachazo:2004tm}. The key simplification
of BCFW recursion relations is that by a proper momentum deformation of
two external particles, we have reduced the multi-variable complex
analysis to the single-variable complex analysis.

Using the BCFW recursion relations, we can get very compact analytical
expressions of tree-level amplitudes at the price of introducing the
spurious-pole structure. Although each term may contain these spurious poles,
they will be canceled out after the sum of all terms, since they are
not physical. Nevertheless they are crucial for simple and compact
expressions and  have very beautiful geometrical picture as shown in
\cite{Hodges:2009hk, ArkaniHamed:2009dn, Hodges:2010kq,
ArkaniHamed:2010gg}.

At the one-loop level, although the integrand is still rational
functions of external momenta (i.e., there is only the single-pole
structure), its integral will produce the branch cut
structure\footnote{Although the branch cut can be chosen
arbitrarily, the starting point (branch-point)  has a definite
physical meaning. }. The location of singularities of one-loop
results can be determined by Landau equations \cite{S-matrix} and
these singularities can be divided into the first-type and second-type.
Based on these analytic structures, a reduction method
\cite{Bern:1992, Ellis:2007, Passarino:1978jh,Neerven:1984} has been
proposed and becomes the standard method for one-loop amplitudes.
The reduction tells us that a general one-loop scatting amplitude
may be expanded in terms of master integrals  with rational
coefficients. This expansion has split the branch cut structure into
the master integrals, while rational coefficients contain the
information of locations of singularities. Since the master
integrals are relatively well understood \cite{Bern:1992,
Ellis:2007}, the one-loop calculation is reduced to the computation
of these coefficients of the master integrals. Based on the rational
structure of one-loop integrands, the very powerful OPP method
\cite{Ossola:2006us, Ellis:2007br} and Forde's methods
\cite{Forde:2007mi, Badger:2008cm} have been proposed. Based on the
branch cut structure, a unitarity cut method was initiated by Bern
{\it et al} \cite{Bern:1994zx, Bern:1994cg} and  was further
generalized first in $4$ dimensions\cite{Britto:2004td,
Britto:2005fm, Britto:2006sj} and then in generalized $D$
dimensions\cite{Anastasiou:2006jv,
Anastasiou:2006gt,Britto:2007tt,Britto:2008vq,Feng:2008gy,Mastrolia:2009dr}.

Although these well known analytic structures of one-loop amplitudes
have led us to very powerful computation methods such as OPP method
and the unitarity cut method, there are still some problems regarding
the analytic structure to be solved. The first problem is that
although from Landau equations we can determine the location of
singularities,  the degree of singularities is not fully discussed. The
information of degree is very important both for theoretical study
and practical calculations, such as the rational part of one-loop
amplitudes. The rational part will appear if the expansion master
basis is the pure 4D scalar box, scalar triangle, scalar bubble or
scalar tadpole. As carefully discussed  in
\cite{Bern:2005hs,Bern:2005ji, Bern:2005cq}, the rational part of
one-loop amplitudes contains  the double pole like ${1\over
\Spaa{4|5}^2}$ for $A_5(1^-,2^+,3^+,4^+,5^+)$ for the gauge theory. Thus
to be able to use the recursion relation to calculate, we need to
determine the degree of poles as well as their residues. Besides the
degree of poles, we will meet the structure like ${\Spbb{a|b}\over
\Spaa{a|b}}$ which is just a phase in the physical region, but a
true pole in the general complex plane. The third related analytic
property we would like to understand is the factorization property
\cite{Bern:1995ix}when $(K^2-M^2)\to 0$. Unlike the tree-level amplitudes with the
factorization property $A_n^{tree}\to A_L^{tree} A_R^{tree}$, for
one-loop amplitudes, we have $A^{1-loop}\to
A_L^{tree}A_R^{1-loop}+A_L^{1-loop} A_R^{tree}+ {\cal F}$ and there
is still no general theory for the structure of the extra term
${\cal F}$.

Above discussions are main motivations of our current investigation.
To learn more about the analytical structure of one-loop amplitudes,
it is always very helpful if we can find explicit expressions for one-loop
amplitudes. With recent developments, especially the unitarity cut methods,
 now we are able to do so. In  \cite{Britto:2007tt},
analytic expressions for various coefficients of master integrals have been given
in the spinor forms. However from the spinor forms it  is
hard to read out analytic properties, thus the first step toward a better understanding
is to translate the spinor form into the Lorentz-invariant contraction
of external momenta.

After the coefficients are written in manifestly Lorentz-invariant
forms, many analytic properties become obvious. The old analysis
made in \cite{S-matrix} tells that singularities can be divided into
the first-type and the second-type.  The first-type singularities
are fully determined by the dual diagrams and all occur in the
scalar bases (master integrals) which are well understood. For the
second-type singularities, our results seem to indicate that they
appear only in coefficients of bases. Although it is well known that
the second-type singularities depend on the dimensionality of
space-time,  the spins of particles, and the details of the their
interactions \cite{Second}, their dependence is not fully discussed.
The main reason is that given the topology of Feynman diagrams, the
denominator of integrand is determined while the numerator can be
arbitrary. In fact, it is the detail of the numerator that defines the
theory under consideration. With our general explicit results
presented in this paper, we are able to have a better
understanding of dependence of the analytic structure on
numerators.
 For example, the degree of second-type
singularities appearing in triangle and bubble coefficients will correlate
with the degree of the internal momentum $\W\ell$ in the numerator. This is one new result
coming out from our analysis.

Having analytic expressions for coefficients will open a door for other
analysis. Just like the tree-level BCFW recursion relations, we can take
pair of momenta to make the deformation, thus reduce the multi-variable complex
analysis to the single-variable complex analysis. The hope is that with these
understanding, one can find similar recursion relations for
one-loop coefficients. Furthermore, the Lorentz-invariant forms of
one-loop coefficients are also the preparations for two-loop
calculations\footnote{Recent new techniques for amplitude
calculations at two- and higher-loop can be found in
\cite{Mastolia:2012ts,Badger:2012dp,Zhang:2012ce,Gluza:2010ts, Kosower:2011ty}.}
using the unitarity cut method.

We must emphasize that although our final goal is to address above problems,
the result in this paper is just the first step toward this goal and there are
still a lot more to be done in future. Thus our results are purely theoretical
orientated and there is nothing to do with improving the efficiency of current
powerful methods for one-loop calculations.

Having above motivations in the mind, in this paper, we show how to
transform the spinor form of one-loop coefficients into the
Lorentz-invariant forms. The evaluation is done within the spinor formalism \cite{Spinor:1981},
reviewed in \cite{Dixon:2005}.
Using these Lorentz invariant forms, we further
discuss the analytic structures of  coefficients, with some
clarifications and interpretations using the S-matrix theory
\cite{S-matrix}.

The outline of this paper is as follows.  In section 2, we give a
brief review of the $D$-dimensional unitarity cut method and the
derivation of the one-loop coefficients. At same time, some
conventions and notations are set up. In section 3, some knowledge
about Landau equations and singularities of S-matrix programm are
reviewed. This section is important to understand our results. In
section 4, we transform the spinor forms of pentagon and box
coefficients into the Lorentz-invariant forms. To enable readers to grasp
main points of our calculations, we have summarized the result in the first subsection 4.1
and leave the details of derivations in later subsections.
 We do similar organizations
in section 5 and 6 for triangle and bubble coefficients
respectively.
Finally in the conclusion section, we  summarize main points of the paper
and have several discussions regarding various possible further applications
and clarifications.
In Appendix A, an typical formula, which is  important in the process from
the spinor form to the Lorentz-invariant form, is given with the proof.

\section{Setup}
In this section, we briefly review the $(4-2\eps)$-dimensional
Unitarity method \cite{Anastasiou:2006jv, Anastasiou:2006gt} and the
derivation of one-loop coefficients
\cite{Britto:2007tt,Britto:2008vq,Feng:2008gy,Britto:2008FY}, which
are the foundation of our work. In this process, we also set up some
key conventions and notations for latter calculations. Here we use
the QCD convention for the square bracket $[i~j]$, so that $2
k_i\cdot k_j = \vev{i~j}[j~i]$.
\subsection{Unitarity cut method}
The unitarity cut of a one-loop amplitude is its discontinuity
across a branch cut in a kinematic region selecting a particular
momentum channel. By denoting the momentum vector across the cut as
$K$, the discontinuity for a double cut can be written as
\bea \label{cut-integral}
C&=&-i(4\pi)^{D/2}\int {d^Dp\over (2\pi)^D}\delta^{(+)}(p^2-M_1^2)\delta^{(+)}((p-K)^2-M_2^2) ~ {\cal T}(p),
\eea
where
\bea {\cal T}(p) = A_{\textrm{Left}}^{\textrm{tree}}(p)\times
A_{\textrm{Right}}^{\textrm{tree}}(p).
 \eea
The ${\cal T}(p)$ can be calculated by any method, for example
Feynman diagrams, off-shell recursion relations \cite{Berends:1987me}
or BCFW on-shell recursion relations \cite{Britto:2004ap,
Britto:2005fq}.

The "unitarity cut method" combines the unitarity cuts with the
familiar PV-reduction method \cite{Passarino:1978jh}. PV-reduction
tells us that any one-loop amplitude can be expanded in master
integrals $I_i$
\bea \label{A-exp} A^{1-loop}=\sum_i c_i I_i. \eea
The master integrals in $(4-2\eps)$-dimensions are tadpoles,
bubbles, triangles, boxes and pentagons\footnote{For massless
external particles, tadpoles do not show up. If we constraint to
pure 4D case, pentagons will not show up, but rational terms appear.
}. It is worth to emphasize that the basis includes the dimensional-shifted
basis \cite{Bern:1996ja}. More explicitly, if we split the $(4-2\eps)$-dimensional internal momentum
$p=p_4+\mu$ with $p_4$ the component in the 4D and $\mu$ the component in
the $(-2\eps)$-dimension, then the basis include
the one like
\bea I_m^D[(\mu^2)^r] \equiv i (-)^{m+1} (4\pi)^{D/2} \int
{d^4p_4\over (2\pi)^4} {d^{-2\eps} \mu\over (2\pi)^{-2\eps}}
{(\mu^2)^r \over
(p^2-\mu^2)...((p-\sum_{i=1}^{m-1}K_{i})^2-\mu^2)}~~\label{D-basis}\eea
$I_m^D[(\mu^2)^r]$ can be translated into the scalar basis with dimensional shifting
as
\bea
I_m^{D=4-2\eps}[(\mu^2)^r]=-\eps(1-\eps)...(r-1-\eps)I_m^{D=4+2r-2\eps}[1]~~~\label{D-shift}\eea
It is, in fact, coefficients of these dimensional shifted bases
producing the rational part of one-loop amplitudes mentioned in the
introduction. To see it, let us notice that, for example
\cite{Heinrich:2010ax},
\bea \int d^{4-2\eps} p {\mu^2\over D_i D_j} & = & -{i\pi^2\over
2}\left[m_i^2+ m_j^2-{(p_i-p_j)^2\over 3}\right]+{\cal
O}(\eps)\nn
\int d^{4-2\eps} p {\mu^2\over D_i D_j D_k} & = & -{i\pi^2\over
2}+{\cal O}(\eps)\nn
\int d^{4-2\eps} p {\mu^4\over D_i D_j D_k D_l} & = & -{i\pi^2\over
6}+{\cal O}(\eps)\nn
\int d^{4-2\eps} p {\mu^2 q^\mu\over D_i D_j D_k} & = &
+{i\pi^2\over 6}(p_i+p_j+p_k)^\mu+{\cal O}(\eps)\nn
\int d^{4-2\eps} p {\mu^2 q^\mu q^\nu\over D_i D_j D_k D_l} & = &
-{i\pi^2\over 12}g^{\mu\nu}+{\cal
O}(\eps)~~~\label{One-rational}\eea
where $D_i\equiv (p+p_i)^2-m_i^2$.
%
%

In the unitarity cut method, we derive the coefficient by
performing unitarity cuts on  both sides of Eq.(\ref{A-exp}):
\bea
\Delta A^{1-loop}=\sum_i c_i \Delta I_i.
\eea
If we can calculate the left-hand side, by comparison, we can read
out the wanted coefficients $c_i$ at the right-hand side.

The $(4-2\epsilon)$-dimensional Lorentz-invariant phase-space (LIPS)
of a double cut is defined by inserting two $\delta$-functions
representing the cut conditions:
\bea \label{LIPS}
  \int {d^{4-2\eps}p}~
\delta^{(+)}(p^2-M_1^2) \delta^{(+)}((p-K)^2-M_2^2)
\eea
where $K$ is the momentum flowing along the double-cut.
To simplify LIPS, we can decompose $(4-2\epsilon)$-dimensional
momentum $p$ as
\bea p = \W\ell + \vec\mu ; \qquad \int d^{4-2\eps} p = \int
d^{-2\eps}\mu  \int d^4 \W \ell \ , \quad \label{Wl-def}\eea
where $\W\ell$ belongs to 4-dimensional part and $\vec\mu$,
$(-2\eps)$-dimensional part. The 4D part momentum $\W\ell$ can be
further decomposed as
\bea \W\ell = \ell + z K, \qquad \ell^2 = 0 ; \qquad \int d^4\W\ell
= \int dz~d^4\ell~ \delta^{(+)}(\ell^2)~(2\ell\cdot K)  ~,\qquad
\eea
where $K$ is the pure 4D cut momentum and $\ell$ is pure 4D massless
momentum, which can be expressed with spinor variables as
\bea \ell=t P, \quad P=
\ket{\ell}\bket{\ell} ;\qquad  \int d^4\ell \ \delta^{(+)}(\ell^2)
&=& \int \vev{\ell~d\ell}[\ell~d\ell] \int t ~dt . \eea

Under this decomposing procedure, Eq.(\ref{LIPS}) becomes
\bea \label{UPS-1} \int d^{4-2\eps} \Phi
 &=&\frac{(4\pi)^\eps}{\Gamma(-\eps)}~\int d\mu^2 ~(\mu^2)^{-1-\eps}
\int {d^4\W\ell}~ \delta^{(+)}(\W\ell^2-\mu^2-M_1^2) ~\delta^{(+)}((\W\ell-K)^2-\mu^2-M_2^2) \nn
 &=&\frac{(4\pi)^\eps}{\Gamma(-\eps)}
~\int d\mu^2 ~(\mu^2)^{-1-\eps} \int \vev{\ell~d\ell}[\ell~d\ell]
{(1-2z)K^2+M_1^2-M_2^2\over\Spab{\ell|K|\ell}^2}\eea
where we have used $\delta$-functions to solve parameters $t$ and
$z$ as
\bea t={(1-2z)K^2+M_1^2-M_2^2\over\Spab{\ell|K|\ell}},\qquad z = {
(K^2+M_1^2-M_2^2)- \sqrt{\Delta[K, M_1, M_2]- 4 K^2\mu^2}\over 2
K^2}, \eea
with the definition
\bea \Delta[K,M_1,M_2]& \equiv & (K^2-M_1^2-M_2^2)^2-4M_1^2 M_2^2
\nn
& = & -4 M_1^2 M_2^2 \left| \begin{array}{cc} 1 &
-{K^2-M_1^2-M_2^2\over 2M_1 M_2}\\ -{K^2-M_1^2-M_2^2\over 2M_1 M_2}
& 1 \end{array}\right|.~~\label{Bubble-Landau}\eea
For convenience, the $\mu^2$-integral measure can be redefined as
\bean \int d\mu^2 (\mu^2)^{-1-\eps} = \left( {
\Delta[K,M_1,M_2]\over 4 K^2}\right)^{-\eps} \int_0^1 du \
u^{-1-\eps}, \eean
where the relation between $u$ and $\mu^2$ is
given by
\bea u\equiv {4 K^2\mu^2 \over \Delta[K,M_1,M_2]}. \quad
\label{u-def}\eea
Using the new variable $u$, we can rewrite $z,t$ as
\bea z ={ \a -\b \sqrt{1-u}\over 2}, \qquad t = {\b\sqrt{1-u} ~ {K^2
\over \Spab{\ell|K|\ell}}} , \quad \label{zt-sol-u} \eea
where
\bea \a = { K^2+M_1^2-M_2^2\over K^2}, \qquad \b={\sqrt{\Delta[K,
M_1, M_2]}\over K^2}. \quad \label{ab-def}\eea

Putting all together, the cut integral Eq.(\ref{cut-integral}) is transformed to
\bea \label{phase-space} C&=& \frac{(4\pi)^\eps}{i \pi^{D/2}
\Gamma(-\eps)}~\left( { \Delta[K,M_1,M_2]\over 4 K^2}\right)^{-\eps}
\int_0^1 du \ u^{-1-\eps} \nn &&\times\int
\vev{\ell~d\ell}[\ell~d\ell] ~\b\sqrt{1-u} ~ {K^2 \over
\Spab{\ell|K|\ell}^2} {\cal T}(p). \eea
where ${\cal T}(p)$ should be  interpreted as
\bea {\cal T}(p) = {\cal T}(\W\ell~,\mu^2) = {\cal
T}(tP+zK ~,\mu^2) = {\cal
T}(\ket{\ell},\bket{\ell},\mu^2) , \quad \eea
with
\bea \W\ell=tP+zK={K^2\over \Spab{\ell|K|\ell}}
\left[\beta\left(P-{K\cdot P\over
K^2}K\right)+ \alpha{K\cdot P\over K^2}K \right]. \eea
\subsection{Input }
For standard quantum field theory\footnote{For non-local theories or
some effective theories, the assumption of input in (\ref{Input}) is
not right.}, ${\cal T}(p)$ is always a sum of following terms\footnote{Since all external
momenta are in pure 4D, the contributions of $p$ in $(4-2\eps)$-dimension can only
have either $\mu^2$-combination or $\W\ell\cdot R_j$-combination. }
\bea {\cal T}(\W\ell)= { \prod _{j=1}^{n+k}(2\W\ell\cdot R_j) \over
\prod_{i=1}^{k}((\W\ell-K_i)^2-m_i^2-\mu^2)}.~~~\label{Input} \eea
where $R_j$ is a generic momentum coming from the Feynman rule (such as polarization vectors)
for general theory. The number of propagators is given by $k$ (the two cut propagators
are not included), thus to have triangles in
the expansion, we need to have $k\geq 1$. To have boxes, $k\geq 2$ and
pentagons $k\geq 3$.  The degree of $\W\ell$ in numerator is given by
$n+k$ where $n$ is a integer, and we require $n+k\geq 0$ only.
For the renormalizable theory, we have $n\leq 2$. However, in this paper,  our discussion
adapts to an arbitrary $n$, such as gravity theory.

If we define
\bea \W R =\sum_{j=1}^{n+k}x_j R_j, \eea then $\prod
_{j=1}^{n+k}(2\W\ell\cdot R_j)$ is just the   $\prod_j^{n+k}x_j$-
component after expanding $ (2\W\ell\cdot \W R)^{n+k}$. So, for
simplicity of our general discussions, we just need take the following
form as the input:
\bea {\cal T}(\W\ell)= { (2\W\ell\cdot \W R)^{n+k}\over
\prod_{i=1}^{k} ((\W\ell-K_i)^2-m_i^2-\mu^2)}~~~\label{Input-1}. \eea

According to the simplified phase-space integration
Eq.(\ref{phase-space}),  the cut integral can be written as
\bea
 C&=&\frac{(4\pi)^\eps}{i \pi^{D/2}\Gamma(-\eps)}~\left( {
\Delta[K,M_1,M_2]\over 4 K^2}\right)^{-\eps} \int_0^1 du \
u^{-1-\eps}\nn &&\times \int
\vev{\ell~d\ell}[\ell~d\ell]~\b\sqrt{1-u}~{ (K^2)^{n+1}\over
\gb{\ell|K|\ell}^{n+2}} {\gb{\ell|R|\ell}^{n+k}\over \prod_{i=1}^k
\gb{\ell|Q_i|\ell}}.~~\label{cut-integral-1} \eea
In the above equation,
\bea
R=\beta(\sqrt{1-u})r+\alpha_R K, ~~~~~~Q_i=\beta(\sqrt{1-u})q_i+\alpha_i K
\eea
where
\bea r&=&\wt R - {\wt R \cdot K\over
K^2}K,~~~~~~~\alpha_R=\alpha{\wt R \cdot K\over K^2}\nn
q_i&=&K_i-{K_i \cdot K\over K^2}K,~~~~~\alpha_i=\alpha{K_i\cdot
K\over K^2}-{K_i^2+M_1^2-m_i^2\over K^2}~. ~~~\label{r-q-def}\eea
For the integrand
\bea I & = &  {(K^2)^{n+1}\over \gb{\ell|K|\ell}^{n+2}} { \gb{\ell|R
|\ell}^{n+k} \over \prod_{i=1}^k \gb{\ell|Q_i
|\ell}},~~~\label{Integrand} \eea
based on spinor formalism, it can be split into
\bea
I=\sum_{i=1}^k F_i(\ell) {1\over
\gb{\ell|K|\ell}\gb{\ell|Q_i|\ell}} +\sum_{q=0}^{n} G_q(\ell)
{\gb{\ell|R|\ell}^q\over \gb{\ell|K|\ell}^{q+2}}~, \qquad \label{I-split}
\eea
where
\bea F_i(\ell) &=& {(K^2)^{n+1}\over \vev{\ell|K Q_i|\ell}^{n+1}}{ \vev{\ell|R Q_i|\ell}^{n+k}\over
\prod_{t=1,t\neq i}^k \vev{\ell|Q_t Q_i |\ell}}~, \qquad\label{Fi}\\
 G_q(\ell)&=&\sum_{i=1}^k{(K^2)^{n+1}\Spaa{\ell|R Q_i|\ell}^{n-q+k-1}\Spaa{\ell|KR|\ell}\over \Spaa{\ell|KQ_i|\ell}^{n-q+1}\prod_{t=1,t\neq i}^k\Spaa{\ell|Q_tQ_i|\ell}}~. \qquad\label{Gq}
\eea

The $F_i$ term contributes to pentagon, boxes and triangles, while
the $G_q$ term,  bubbles. Substituting the splitting result into
Eq.(\ref{cut-integral-1}), and taking the residues of different
poles, we can get coefficients of various master integrals.

\subsection{Summary of coefficients}

Now we list the coefficients of different master integrals. The
pentagon and box coefficients are given by
\bea C[Q_i,Q_j,K]={(K^2)^{n+2}\over
2}\left({\Spab{P_{ij,1}|R|P_{ij,2}}^{n+k}\over
\Spab{P_{ij,1}|K|P_{ij,2}}^{n+2}\prod_{t=1,t\neq
i,j}^k\Spab{P_{ij,1}|Q_t|P_{ij,2}}}+\{P_{ij,1}\leftrightarrow
P_{ij,2}\}\right)~~~\label{Box-pc} \eea
where $P_{ij,1}$ and $P_{ij,2}$ are two massless momenta constructed
from $Q_i$ and $Q_j$ ($i\leq j$).  More explicitly, if both $Q_i, Q_j$ are
massless, then we can set $Q_i=P_{ij,1}$ and $Q_j=P_{ij,2}$. If one of them
is not massless, for example, $Q_i^2\neq 0$, we can construct
\bea P_{ij}=(Q_j+x Q_i)~~~~\label{P1P2}.\eea
The condition $P_{ij}^2=0$ leads to following two solutions of $x$
\bea x_{1,2} & = & { -2 Q_i\cdot Q_j\pm \sqrt{ (2 Q_i\cdot Q_j)^2-4
Q_i^2 Q_j^2 }~~~ \over 2 Q_i^2}\label{Pole-construct}, \eea
thus we have constructed two massless momenta. Formula \eref{Box-pc}
makes sense when and only when $k\geq 2$. Furthermore, if $n\leq -3$
(noticing that we need to have $n+k\geq 0$), there is only pentagon
coefficients.

The triangle coefficient is given by
\bea C[Q_i,K]&=&{(K^2)^{n+1}\over 2}{1\over
(\sqrt{\Delta})^{n+1}}{1\over (n+1)!
\Spaa{P_{i,1}~P_{i,2}}^{n+1}}\nn
 &&\times {d^{n+1}\over d
\tau^{n+1}}\left.\left(\left.{\Spaa{\ell|R Q_i|\ell}^{n+k}\over
\prod_{t=1,t\neq i}^k\Spaa{\ell|Q_t Q_i|\ell}}\right|_{\ell\to
P_{i,1}-\tau P_{i,2}} +\{P_{i,1}\leftrightarrow
P_{i,2}\}\right)\right|_{\tau\to 0}~~\label{Tri-coeff-1} \eea
where $P_{i,1}$ and $P_{i,2}$ are two massless momentum constructed from
$K$ and $Q_i$. Formula \eref{Tri-coeff-1} makes sense when and only
when $k\geq 1$ and $n\geq -1$.

The coefficient of the bubble is the sum of the residues of the poles from the following expression:
\bea B&=&\sum_{i=1}^k\sum_{q=0}^{n}{-(K^2)^{n+1}\Spaa{\ell|R
Q_i|\ell}^{n-q+k-1}\over
\Spaa{\ell|KQ_i|\ell}^{n-q+1}\prod_{t=1,t\neq
i}^k\Spaa{\ell|Q_tQ_i|\ell}} {1\over
q+1}{\Spab{\ell|R|\ell}^{q+1}\over \Spab{\ell|K|\ell}^{q+1}}
~~~\label{bubble-coeff-1}\eea
where poles are given by factors $\Spaa{\ell|KQ_i|\ell}$ and
$\Spaa{\ell|Q_tQ_i|\ell}$. Formula \eref{bubble-coeff-1} makes sense
when and only when $k\geq 0$ and $n\geq 0$.

\subsection{Notations}

For convenience, we give notations we will adopt in the paper. First we define the
following determinant
\bea  G\left( \begin{array}{cccc} p_1 & p_2 & ... & p_k \\
q_1 & q_2 & ... & q_k \end{array}\right)={\rm det} \left( p_i\cdot q_j\right)_{k\times k}.~~~
\label{Gram-def}\eea
If $q_i=p_i$ we  write
\bea G(p_1,p_2,...,p_k)\equiv G\left( \begin{array}{cccc} p_1 & p_2 & ... & p_k \\
p_1 & p_2 & ... & p_k \end{array}\right).~~~
\label{Gram-def-1}\eea
If $q_i=p_i$ for $i=1,...,k-1$ in \eref{Gram-def}, for short we can write it as
\bea (p_k|q_k)|_{p_1,..,p_{k-1}}= G\left( \begin{array}{ccccc} p_1 & p_2 & ... & p_{k-1} & p_k \\
p_1 & p_2 & ... & p_{k-1} & q_k\end{array}\right).~~~
\label{Gram-def-2}\eea
If the meaning of $p_1,...,p_{k-1}$ is unambiguous, we can even
write it as $(p_k|q_k)$.

Second we define other determinants which are related to the Gram
determinant, but depend on the masses of propagators. They are
\bea  N^{(K_i,K_j,K;\wt R)}& = &-\det\left(
                        \begin{array}{cccc}
                          0              & K^2+M_1^2-M_2^2 & K_i^2+M_1^2-m_i^2  & K_j^2+M_1^2-m_j^2 \\
                          \wt R\cdot K   & K^2             & K_i\cdot K          & K_j\cdot K \\
                          \wt R\cdot K_i & K\cdot K_i      & K_i^2               & K_j\cdot K_i  \\
                          \wt R\cdot K_j & K\cdot K_j^2    & K_i\cdot K_j        & K_j^2 \\
                        \end{array}
                      \right)~~~\label{N-4var}\eea
with $4$ parameters and the structurally similar
\bea &&N^{(K_i,K_j,K_t,K;\wt R)}=\nn &&-\det\left(
                          \begin{array}{ccccc}
                            0            & K^2+M_1^2-M_2^2&K_i^2+M_1^2-m_i^2&K_j^2+M_1^2-m_j^2&K_t^2+M_1^2-m_t^2\\
                            \wt R\cdot K&K^2 & K_i\cdot K & K_j\cdot K & K_t\cdot K \\
                            \wt R\cdot K_i&K_i\cdot K  & K_i^2 &K_i\cdot K_j & K_t\cdot K_i\\
                            \wt R\cdot K_j&K_j\cdot K  & K_i\cdot K_j  & K_j^2 & K_t\cdot K_j  \\
                            \wt R\cdot K_t&K_t\cdot K  & K_i\cdot K_t  & K_j\cdot K_t  & K_t^2\\
                          \end{array}
                        \right).~~\label{Pen-N}
\eea
with $5$ parameters. Another one is
 \bea &&D^{(K_i,K_j,K_t,K_s,K)}=\nn &&\det\left(
                          \begin{array}{ccccc}
                           K ^2+M_1^2-M_2^2&K_i ^2+M_1^2-m_i^2&K_j ^2+M_1^2-m_j^2&K_t ^2+M_1^2-m_t^2&K_s ^2+M_1^2-m_s^2\\
                            K\cdot K_i&K_i^2 & K_i\cdot K_j & K_i\cdot K_t & K_i\cdot K_s \\
                            K\cdot K_j&K_i\cdot K_j  & K_j^2 &K_t\cdot K_j & K_j\cdot K_s\\
                            K\cdot K_t&K_i\cdot K_t  & K_t\cdot K_j  & K_t^2 & K_t\cdot K_s  \\
                            K\cdot K_s&K_i\cdot K_s  & K_j\cdot K_s  & K_t\cdot K_s  & K_s^2\\
                          \end{array}
                        \right),~~~~\label{Hex-pen}
\eea
which is related to reducing the hexagon to the pentagon.

\section{Singularities}

One main motivation of our calculations it to discuss the analytic
structure of coefficients of master integrals. For this purpose, in
this section, we will review some backgrounds coming from the study
of S-matrix program \cite{S-matrix,Second}. The main point is that
locations of all possible singularities of a Feynman integral can be
determined, in principle, by the Landau equations. These
singularities can be divided into two types: the  first-type and the
second-type. However, as we have mentioned in the introduction, the degree
of singularities, especially the second-type singularities, has not been
 discussed in \cite{S-matrix,Second}.

\subsection{Landau equations}
To start, let us notice that  apart from constant multiplicative
factors, after Feynman parametrization, the general Feynman integral
takes the form
\bea
I&=&\int{\nu({q})\delta(\sum_i \alpha_i-1)\left(\prod\limits_{i=1}^Nd\alpha_i\right)\left(\prod\limits_{j=1}^N d^n{k_j}\right)\over \psi^N},~~~\label{Fey-1}
\eea
with $\psi$ defined by
\bea
\psi({p},{k},\alpha)&=&\sum_{i=1}^N \alpha_i(q_i^2-m_i^2).
\eea
Here $N$, $l$ are, respectively, the numbers of the internal lines and the independent loops of the corresponding graph.
$\alpha_i$, ${q_i}$, $m_i$ are, respectively, the Feynman integration parameter, the momentum, and the mass associated with the $i$th line.
$\nu({q})$ is a polynomial which involves the spins of the participating particles and the details of their interactions.
$n$ is the dimensionality of Lorentz space.

The four momentum $q_i$ in any internal line is a linear function of the circulating momenta $k$ and the external momenta $p$.
Therefore the quadratic form $\psi({p},{k},\alpha)$ can be written as
\bea \psi({p},{k},\alpha)&=&\sum_{i,j=1}^la_{i,j}k_i
k_j+\sum_{j=1}^lb_j k_j +c\nn &=&\bm{k^T\cdot Ak-2k^T\cdot
Bp}+(\bm{p^T\cdot \Gamma p}-\sigma),~~~\label{psi} \eea
where
\bea
\sigma=\sum_i \alpha_i m_i^2.~~~\label{sigma} \eea

Here, $\bm{A,B,\Gamma }$ are respectively $l\times l, l\times (E-1),
(E-1)\times (E-1)$ matrices, whose elements are linear in $\alpha$.
$E$ denotes the number of external lines of the diagram. $k$ and $p$
are column vectors in the spaces of the matrices and their elements
are themselves Lorentz four-vectors.

For most discussions,  $\nu(q)=1$ for (\ref{Fey-1}) has been assumed. It is
enough for the location of singularities.
Performing the integration over $k$ in Eq.(\ref{Fey-1}), we can get
\bea
I&=&\int{ C^{N-(1/2)n(l+1)}\delta(\sum_i \alpha_i-1)\left
(\prod\limits_{i=1}^Nd\alpha_i\right)\over D^{N-(1/2)nl}},~~~\label{Fey-2}
\eea
where
\bea C={\rm det}(\bm {A}),~~~D&=&\bm{-(Bp)^T\cdot X(Bp)+(p^T\cdot \Gamma
p}-\sigma)C, \eea
with $X={\rm adj}(A)$\footnote{$X$ is always well defined even ${\rm
det}(A)=0$. If ${\rm det}(A)\neq 0$, we have  $X=A^{-1}C$.} and
$\sigma$ defined by Eq.(\ref{sigma}). $C$ is of degree $l$ in the
$\a$ and $D$, of degree $(l+1)$. According to the generalized
Hadamard lemma, the necessary conditions for a singularity of $I$
are, using the representation Eq.(\ref{Fey-2}),
\bea {\rm Form~I}:~~~~\alpha_i {\partial D\over \partial
\alpha_i}=0,~~\hbox{for each $i$}.~~~\label{Sing-Form-1} \eea
If we use the representation (\ref{Fey-1}), the Landau equations
will be given by
\bea {\rm Form~II}:~~~~\left\{
\begin{array}{ll}\alpha_i(q_i^2-m_i^2)=0, &~~\hbox{for each propagator
$i$}\\ \sum_j \alpha_i q_i=0, & ~~\hbox{for each loop running by
loop-momentum $k_j$}
\end{array}\right.~~~\label{Sing-Form-2}\eea
In both forms (\ref{Sing-Form-1}) and (\ref{Sing-Form-2}), solution
with $\a_i=0$ corresponds to pinch the corresponding propagators, so
for example, a box diagram will reduce to a triangle diagram. The
singularity of a given graph with no $\a_i=0$ (i.e., all propagators
are on the mass shell) is called the "leading singularity".

A connection between these two forms can be found by noticing that
 an alternative expression for $D$ is given by
\bea D=CD', \eea
where $D'$ is the result of eliminating $k$ from
$\psi$ by means of the equations
\bea {\partial \psi\over
\partial k_j}=0, ~~~\hbox{for each $j$}.~~\label{p-psi} \eea
In the notation of Eq.(\ref{psi}), these equations are
\bea
\bm{Ak=Bp}. ~~~\label{Landau-rel} \eea
Together with Eq.(\ref{psi}) and Eq.(\ref{p-psi}), we obtain the
Landau equations (\ref{Sing-Form-2}).

\bea \sum_j \alpha_i q_i=0, ~~~\hbox{for each
$j$},~~\label{q-linear} \eea and \bea
\alpha_i(q_i^2-m_i^2)=0,~~\hbox{for each $i$},~~\label{alpha-q} \eea
where $\sum_j$ in Eq.(\ref{q-linear}) denotes summation round the
$j$th closed loop of the diagram.

\subsection{Singularities of the first type}

The Landau equations are usually too complicated to solve
algebraically. So a geometrical method, which is the so called dual
diagram, have been  introduced. The dual diagram is vector diagram
for internal and external momenta. From dual diagrams we can read
out the Landau surface where singularities of the first type may locate. For
example, for bubble diagram, $\Delta[K,M_1,M_2]$ in
(\ref{Bubble-Landau}) is nothing, but exactly the Landau surface.
From this surface, we can find the location of singularities is
$K^2= (M_1\pm M_2)^2$. The Landau surface of triangle is given by
\bea \Sigma_{tri}=\left| \begin{array}{ccc} 1 & - y_{12} & -
y_{13}\\ -y_{21} & 1 & -y_{23} \\ -y_{31} & -y_{32} & 1
\end{array}\right|,~~~\label{tri-1}\eea
where $y_{ij}=y_{ji}={P_k^2-m_i^2-m_j^2\over 2 m_i m_j}$ with
$(i,j,k)$ a permutation of $(1,2,3)$. The $m_i$ is the mass of
the propagator $q_i$ and $P_i$ is the external momentum at the vertex $i$
opposite to the propagator $q_i$. For the box diagram, let us denote
external momenta clockwise as $P_i^2=M_i^2$, $i=1,2,3,4$ and
internal propagators clockwise as $q_i$ with mass $m_i$ ( $q_{i-1},
q_i$ and $P_i$ meet at the same vertex), then the Landau surface is
given by
\bea \Sigma_{box}=\left| \begin{array}{cccc} 1 & - y_{12} & - y_{13} & -y_{14}\\
-y_{21} & 1 & -y_{23} &  -y_{24} \\ -y_{31} & -y_{32} & 1 & -y_{34}
\\ -y_{41} & -y_{42} & -y_{43} & 1
\end{array}\right|~~~\label{box-1}\eea
where $y_{ij}=y_{ji}={(q_i-q_j)^2-m_i^2-m_j^2\over 2 m_i m_j}$.

One important point of the Landau surfaces (\ref{tri-1}) and
(\ref{box-1}) of the first-type singularities is that they depend on
masses of inner propagators.

\subsection{Singularities of the second type}

The conventional dual diagrams do not represent all possible
solutions of the Landau equations. The extra solutions are called
the second-type solutions. They correspond to rather special
solutions of the Landau equations. In Eq.(\ref{Landau-rel}), if
$\bm{A}$ is non-singular, ${\bf k}$ will have a unique solution in
terms of the ${\bf p}$ which will exactly correspond to the dual
diagram construction. Hence second-type solutions will have to
correspond to $\bm{A}$ being singular, that is to the condition
\bea
C=\det\bm{A}=0. \eea

Second-type singularities can be divide into two classes, pure
second-type and mixed second-type. The former, which are given by
the Gram determinant equation (\ref{Gram-def})
\bea G(p_1,...,p_{E-1})=\det p_i\cdot p_j=0,~~i,j=1,...,E-1,~~~\label{Gram-d} \eea
where $p_i$ represent any $(E-1)$ of the $E$ external momenta of the
graph. The equation (\ref{Gram-d}) is the condition that there be a
linear combination of the vectors $p_1\ldots p_{E-1}$ equal to zero
or, more generally, equal to a zero-length vector whose scalar
products with $p_1,\ldots,p_{E-1}$ are zero. Detailed analysis
reveals that second-type singularities stem from super pinches at
infinity and correspond to infinite values for some of the
components of the internal momenta in the Feynman graph.

Second-type singularities have some properties. First the curve
given by (\ref{Gram-d}) is {\sl independent of the masses} of the
internal particles. Secondly, the presence of second-type
singularities involves the dimensionality of space,  the spins of
particles, and the details of the their interactions. For example,
for pure scalar theory, i.e., $\nu(q)=1$ in (\ref{Fey-1}), only when
$E<n$ ( $n$ is the dimension of space-time), second-type singularity
exists. This result will be changed if $\nu(q)$ is nontrivial
function.

In a diagram with several loops,  there may be super pinches only
for some of the loop momenta while the others have ordinary pinches
at finite points. These singularities are called {\sl mixed
second-type singularities} and their equations will depend upon the
internal masses of the lines round the loops
 with finite loops. In this paper, we will focus on one-loop
 diagrams, so we will not meet the mixed second-type singularities.


\section{Coefficients of pentagon and box}

Starting from this section, we will present the explicit Lorentz-invariant form of
external momenta for various coefficients of master integrals. Since the transformation
from the spinor form to the Lorentz-invariant form is a little bit complicated, we will summarize
the main results at the beginning of each section and leave the derivation
 in the later part, for  which readers can skip safely if they want.

The pentagon and box coefficients are given by Eq.(\ref{Box-pc}). In
subsection 4.1, we summarize our results and discuss analytic
properties of pentagon and box coefficients derived from our
calculations. In subsection 4.2, we discuss how to separate pentagon
and box coefficients from the single expression (\ref{Box-pc}). Then
we evaluate box coefficients in subsection 4.3. To do so,
we need carry out a typical sum, which is done in Appendix (see
\eref{Sig-Def} and \eref{Sig-E}). The same sum patten appears also
for triangle and bubble coefficients.

\subsection{The summary of main results of current section}

{\bf Pentagon:} First, the Lorentz-invariant forms of the pentagon
coefficients defined by momenta $K_i,K_j,K_t,K$ are
\bea C[Q_i, Q_j, Q_t,K]
&=&\left({{N^{(K_i,K_j,K_t,K;\wt R)}\over
G(K_i,K_j,K_t,K)}}\right)^{n+k}\prod_{w=1,w\neq i,j,t}^k
{G\left( \begin{array}{cccc} K_i & K_j & K_t & K \\
K_i & K_j & K_t & K_\omega  \end{array}\right)\over
D^{(K_i,K_j,K_t,K_\omega,K)}}, ~~~\label{Pen-coeff} \eea
where functions $G$, $N$ and $D$ can be found in \eref{Gram-def-1},
\eref{Pen-N} and \eref{Hex-pen}. From the  expression
(\ref{Pen-coeff}) following analytic properties of pentagon
coefficients can be read out:
\begin{itemize}

\item First the factor $G(K_i,K_j,K_t,K)$ is nothing, but the
second-type singularity intrinsically related to pentagon topology.
Furthermore, its degree is $(n+k)$, which is the degree of $\W\ell$
in numerator of the input \eref{Input}. More explicitly, to have
the pentagon in the expansion, we must have $k\geq 3$ in \eref{Input}
and $n+k\geq 0$. If $n+k=0$, the singularity $G(K_i,K_j,K_t,K)$ does
not appear, but it will be there when $n+k\geq 1$.

\item Secondly, there are singularities given by
$D^{(K_i,K_j,K_t,K_\omega,K)}$. They come from the trivial reduction of the
hexagon topology to the pentagon topology and depend on masses of
propagators. Their dependence of masses is not like that of
first-type singularities  given in (\ref{Bubble-Landau}) for the
bubble, (\ref{tri-1}) for the triangle and (\ref{box-1}) for the
box. In fact, the trivial reduction from the  hexagon to the pentagon  is
intrinsically related to the four-dimensional  space-time. Thus we
guess the appearance of this type of singularities is also related
to the space-time dimension.

\item Thirdly,  $C[Q_i, Q_j, Q_t,K]$ does not contain $u$
at all. In other words, dimensional shifted bases exist only for
box, triangle, bubble and tadpole topologies. This is tightly
related to $(4-2\eps)$-dimensional analysis. In fact, it is well
known that if we do reduction in pure $4$-dimension, the pentagon
will not be a basis at all.

\end{itemize}

It is worth to point out that since the (\ref{Pen-coeff}) is given
by only one term, we will not expect cancelation of any factor in
denominators\footnote{It is worth to emphasize that in this paper we
will not discuss the cancelation of singularities after summing over
 contributions from all bases. For example,  it is very possible that
the singularity like $D^{(K_i,K_j,K_t,K_\omega,K)}$ will be canceled out
if we sum up contributions from all pentagons. }.

~\\

{\bf Box:} The true box coefficients are given by
\bea
C[Q_i,Q_j,K]&=&\sum_{z_1+\ldots+z_k+s=n+2}\sum_{h=0}^s\left(\prod_{t=1,t\neq
i,j}^k {G\left( \begin{array}{cccc} K_i & K_j & K_t & K \\
K_i & K_j & \W R & K
\end{array}\right) \over G(K_i,K_j,K_t,K)}
\left({N^{(K_i,K_j,K_t,K;\wt R)}\over
G(K_i,K_j,K_t,K)}\right)^{z_t}\right)\nn &&\times {s\choose h}{
{N^{(K_i,K_j,K;\wt R)}}^{s-h}T(h)\over
(G(K_i,K_j,K))^s};~~~\hbox{with $h$ even}~~~\label{Box-L-Coef-1}
\eea
 where $T(h)$ is defined in Eq.({\ref{T-def}}) and $G$, $N$ can
  be found in \eref{Gram-def-1},
\eref{Pen-N}. The derivation of \eref{Box-L-Coef-1} is given in
later subsections.

Now  the analytic properties of box coefficients can be read out:
\begin{itemize}

\item First we notice that although there are first-type and
second-type singularities in general, the box coefficient contains
only second-type singularities. The first-type singularity of the
box appears only in the box scalar basis\footnote{It is shown in
\cite{Second} that for 4D, the scalar box basis does not contain the
second-type singularity in {\sl physical sheet}. However,
second-type singularities do appear for scalar triangle and bubble
bases in 4D. The general condition is that $E<D$ where $E$ is the
number of external lines and $D$, dimensions of space-time. It is
worth to emphasize that although second-type singularity of box does
not show up in the physical sheet, it does show up in other sheet.
Thus its understanding is also important for the study of analytic
properties.}.

\item Among all second-type singularities, the one given by
$G(K_i,K_j,K)$ is  intrinsically related to the box topology. Among
all terms of \eref{Box-L-Coef-1}, there is one and only one term
with the highest $s=n+2$ and all other $z_t=0$. Thus this term can
not be canceled by others and the highest degree of singularity
$G(K_i,K_j,K)$ is $(n+2)$.

It is worth to compare the degree of singularities between the
pentagon and the box. The highest degree of the  pentagon
singularity is $(n+k)$ (i.e., the degree of $\W\ell$ in the
numerator) while that of the box singularity is $(n+2)$. Naively,
when we do the reduction, one $\W\ell$ in numerator will cancel one
propagator, thus for box we need to cancel $(k-2)$ propagators, so
the remaining degree of $\W\ell$ in the numerator is
$(n+k)-(k-2)=(n+2)$ as we expect. However, the degree of the
pentagon singularity does not follow the rule. We think the reason
is following. The naive observation is based on the reduction in 4D.
If we do everything in pure 4D, the pentagon will not be a basis as
given in \cite{Bern:1993kr}\footnote{With  formula
\eref{Pen-to-box}, we can change the choice of basis from
$I_5^{D=4-2\eps}[1]$ to $I_5^{D=6-2\eps}[1]$. In this paper, we will
choose $I_5^{D=4-2\eps}[1]$ as our basis to do the PV-reduction. One
of the reason is that with this choice of basis, the pentagon
coefficient will not depend on $u$, thus contributions to one-loop
rational part will be contained completely in the box, triangle and
bubble parts. },
\bea I_5^{D=4-2\eps}[1]=\sum_{i=1}^5 c_i I_{4,i}^{D=4-2\eps}[1]+\eps
I_5^{D=6-2\eps}[1]~~~\label{Pen-to-box}\eea
where $I_5^{D=6-2\eps}[1]$ is finite when $\eps\to 0$. In fact, it
is because we do reduction in $(4-2\eps)$-dimension, the pentagon
becomes a necessary basis. Based on the observation, we think the
naive observation is not applicable to the pentagon topology and
each $\W\ell$ in the numerator does give a contribution to the
degree of the singularity.

\item The appearance of the second-type singularity
$G(K_i,K_j,K_t,K)$ indicates the influence of pentagon topologies,
which will produce the same box topology when pinching one
propagator. Just like the previous item, for  a given singularity
$G(K_i,K_j,K_t,K)$ there is one and only one term inside
\eref{Box-L-Coef-1} with the highest degree $(n+3)$, thus it can not
be canceled by other terms. The highest degree $(n+3)$ of the pole
$G(K_i,K_j,K_t,K)$ can be understood by the naive observation, i.e.,
in the reduction one $\W\ell$ in the numerator will cancel one
propagator. To produce the pentagon topology, we need to get rid of
$(k-3)$ propagators, thus the degree of $\W\ell$ in the numerator
becomes $(n+k)-(k-3)=(n+3)$. Each remaining $\W\ell$ will produce
one $G(K_i,K_j,K_t,K)$ singularity when pinched to box.

\item To see the dimensional shifted basis (which is related to rational
part of one-loop amplitudes, see \eref{D-basis},
\eref{One-rational}), we need to check the $u$-dependence part in the
numerator. From \eref{Box-L-Coef-1}, all $u$-dependence comes from the
factor $T(h)$ and the highest degree of $u$ is $[(n+2)/2]$. It is also
important to notice that each $u$ will be accompanied by a factor
$G(K_i,K_j,K)$ (see Eq.({\ref{T-def}}), which will reduce the
highest degree of the pole $G(K_i,K_j,K)$ for these (rational) terms.

\end{itemize}
%

\subsection{The separation of coefficients of pentagon and box}

Since the expression (\ref{Box-pc}) contains both pentagon and box coefficients,
the first step is to  separate the pentagon coefficient from the box.
This separation has been discussed  in \cite{Britto:2008FY}. However,
since we need write them
more compactly and systematically  and  we
are dealing with the $(4-2\eps)$-dimensional massive case, which is
different from the  massless case in \cite{Britto:2008FY},
 we will give the main steps to write out our results and leave some details
 to be referred to \cite{Britto:2008FY}.

{\bf Expanding numerator:} The first preparation for the separation
is to expand $r$ (see \eref{r-q-def})  in the basis $q_i,q_j,q_t$ as
( remembering $r\cdot K=0$)
\bea \label {r-expa}
r=a_{t}^{(q_i,q_j,q_t;r)}q_t+a_{i}^{(q_i,q_j,q_t;r)}q_i+a_{j}^{(q_i,q_j,q_t;r)}q_j~,
~~~~~\label {r-expa} \eea
which is equal to the expansion  of $\wt R$ in the basis
$K_i,K_j,K_t,K$ because $q_i\cdot K=0$:
\bea \label{R-exp}
\wt R=a_{t}^{(K_i,K_j,K_t,K;\wt R)}K_t+a_{i}^{(K_i,K_j,K_t,K;\wt R)}
K_i+a_{j}^{(K_i,K_j,K_t,K;\wt R)}K_j+a_K^{(K_i,K_j,K_t,K;\wt R)}K.
\eea
By projecting Eq.(\ref{R-exp}) onto the vectorspace orthogonal to $K$, we can easily check:
\bea
a_{\omega}^{(q_i,q_j,q_t;r)}=a_{\omega}^{(K_i,K_j,K_t,K;\wt R)},~~~~\omega=i,j,t.
\eea
The Crammer rule gives the solution of Eq. ({\ref{r-expa}})
\bea
a_\omega^{(q_i,q_j,q_t;r)}={ G\left( \begin{array}{ccccc} K_i & ... & \W R &... & K \\
K_i & ... & K_\omega &... & K \end{array}\right)\over
G(K_i,K_j,K_t,K)},~~~~~\omega=i,j,t~~~\label{a-omega}
\eea
using the notation \eref{Gram-def}.
%
%
The denominator
$G(K_i,K_j,K_t,K)$ is nothing, but the {\sl second-type
singularity} related to the pentagon determined by momenta $K, K_i, K_j,
K_t$. In other words, it can be considered as the "finger print" of
the related pentagon.

Using the expansion Eq. (\ref{r-expa}), we have
\bea
\Spab{P_1|R|P_2}=a_t^{(q_i,q_j,q_t;r)}\Spab{P_1|Q_t|P_2}+\beta^{(q_i,q_j,q_t;r)}\Spab{P_1|K|P_2},
~~\label{R-exp-box} \eea
with
\bea
\beta^{(q_i,q_j,q_t;r)}&=&\alpha_R-\sum_{\omega=i,j,t}a_\omega^{(q_i,q_j,q_t;r)}
\alpha_\omega
={N^{(K_i,K_j,K_t,K;\wt R)}\over K^2
G(K_i,K_j,K_t,K)},~~\label{beta} \eea
where $N$ has been given in \eref{Pen-N}.

~\\

{\bf Separating box from pentagon:} Having explained how to expand
$R$, now we discuss how to separate box from pentagon  in
\eref{Box-pc}. First we give an example like $ {\Spaa{R}^3\over
\Spaa{K}\Spaa{Q_{t_1}} \Spaa{Q_{t_2}}}$\footnote{For simplicity we
have used such short notation $\Spaa{R}=\Spab{P_1|R|P_2}$. By
comparing with \eref{Box-pc}, we hope its meaning is obvious.}.
First using $K,K_i,K_j,K_{t_1}$ to expand one $R$ we will get
\bean {\Spaa{R}^3\over \Spaa{K}\Spaa{Q_{t_1}} \Spaa{Q_{t_2}} }\to
{\Spaa{R}^2\over \Spaa{Q_{t_1}} \Spaa{Q_{t_2}}}+{\Spaa{R}^2\over
\Spaa{K} \Spaa{Q_{t_2}}}\eean
For the first term we expand  $R$ using $K,K_i,K_j,K_{t_1}$ while
for the second term we expand the $R$ using $K,K_i,K_j,K_{t_2}$,
thus we get
\bean \left({\Spaa{R}\over \Spaa{Q_{t_2}}}+{\Spaa{R}\Spaa{K}\over
\Spaa{Q_{t_1}} \Spaa{Q_{t_2}}}\right)+\left( {\Spaa{R}\over
\Spaa{Q_{t_2}}}+{\Spaa{R}\over \Spaa{K} }\right)\eean
Among these four terms, the last one contributes to the box only. For
the first three terms, we  expand the remaining $R$ and arrive
\bean \left(\left[c_1+{\Spaa{K}\over \Spaa{Q_{t_2}}}\right]+ \left[
{\Spaa{K}\over  \Spaa{Q_{t_2}}}+{\Spaa{K}^2\over \Spaa{Q_{t_1}}
\Spaa{Q_{t_2}}}\right] \right)+\left(\left[ c_2+{\Spaa{K}\over
\Spaa{Q_{t_2}}}\right]+{\Spaa{R}\over \Spaa{K} }\right)\eean
The last step is to use $Q_i,Q_j,Q_{t_1},Q_{t_2}$ to expand $K$ for
the fourth term
\bean { \Spab{P_1|K|P_2}\over \Spab{P_1|Q_t|P_2}\Spab{P_1|Q_s|P_2}}=
{-1\over \a_K}\left({ \a_s\over \Spab{P_1|Q_t|P_2}}+ {\a_t\over
\Spab{P_1|Q_s|P_2}}\right)~,\eean
and we arrive
\bean \left(\left[c_1+{\Spaa{K}\over \Spaa{Q_{t_2}}}\right]+ \left[
{\Spaa{K}\over  \Spaa{Q_{t_2}}}+\left({\Spaa{K}\over \Spaa{Q_{t_1}}
}+{\Spaa{K}\over \Spaa{Q_{t_2}} }\right)\right] \right)+\left(\left[
c_2+{\Spaa{K}\over \Spaa{Q_{t_2}}}\right]+{\Spaa{R}\over \Spaa{K}
}\right).\eean
Now we have got the complete splitting. The first, the sixth and
the eighth terms contribute to the box only. The fourth term contributes
to the pentagon $(K,K_i,K_j,K_{t_1})$ only. The second, the third, the
fifth and the seventh terms contribute to the pentagon
$(K,K_i,K_j,K_{t_2})$ only. In our splitting, we have carefully
chosen the way to expand $R$, so that the contribution to the pentagon
$(K,K_i,K_j,K_{t_1})$ appears only once while   the contribution to the
pentagon $(K,K_i,K_j,K_{t_2})$ appears four times. However, it can be
checked that the sum of these four terms does produce only one term.

The above splitting can be generalized to arbitrary $k$ and $n\geq -2$.
First we define
\bea B^1 [n,k]&=&{\Spab{P_1|R|P_2}^{n+k}\over \Spab{P_1|K|P_2}^{n+2}
\prod_{t=1,t\neq i,j}^k\Spab{P_1|Q_t|P_2}},\nn
 B^2
[n,k]&=&\left. B^1 [n,k]\right|_{P_{1}\leftrightarrow P_{2}}. \eea
$B^1[n,k]$ and $B^2[n,k]$ are just the first term and second term respectively in the parenthesis of  Eq.\eref{Box-pc} and $B^2[n,k]$ is obtained from $B^1[n,k]$ by exchanging $P_1\leftrightarrow P_2$.
For the simplest example   $k=3$ we will have  (for example,
$t=3,i=1,j=2$)
\bea B^1[n,3]=\sum_{s=0}^{n+2}C_s[n,3]{\Spab{P_1|R|P_2}^{s}\over
\Spab{P_1|K|P_2}^s}+F[n,3]{\Spab{P_1|K|P_2}\over
\Spab{P_1|Q_3|P_2}}~~~\label {Box-1} \eea
where (\ref{R-exp-box}) has been used. In the above equation, the
first term give the true box coefficient, while the second term, the
pentagon coefficient. The expression of $C_s[n,3]$ and $F[n,3]$ can
be obtained by induction on $n$ as
\bea F[n,3]&=&{\beta^{(q_i,q_j,q_3;r)}}^{n+3},~~~~
C_s[n,3]=a_3^{(q_i,q_j,q_3;r)}{\beta^{(q_i,q_j,q_3;r)}}^{n+2-s}.
~~~\label{Box-C}\eea
For $k\geq 4$, we use the induction on $k$ and get the complete
splitting.

The contribution to the pentagon part has been given in
\cite{Britto:2008FY} (with a generalization to the massive case) and its
evaluation gives \eref{Pen-coeff} while the contribution to box part
is given by the following sum
\bea
C[Q_i,Q_j]^{(n,k)}&=&\sum_{z_1+\ldots+z_k+s=n+2}\left(\prod_{t=1,t\neq
i,j}^k a_t^{(q_i,q_j,q_t;r)}
\left(\beta^{(q_i,q_j,q_t;r)}\right)^{z_t}\right)\left({\Spab{P_1|R|P_2}^{s}\over
\Spab{P_1|K|P_2}^{s}}+\{P_1\leftrightarrow P_2\}\right)~~~~~\label{Box-ind} \eea
where $s,z_t\geq 0$ and in the sum $z_1+\ldots+z_k+s=n+2$, $z_i,
z_j$ should be excluded. This formula is  completely symmetric on
$t$.

\subsection{Evaluation of box coefficients}

Now we need to evaluate \eref{Box-ind}, where the  sum
$\left({\Spab{P_1|R|P_2}^{s}\over \Spab{P_1|K|P_2}^{s}}+\{P_1\to
P_2\}\right)$ appears. This sum is a special case of the typical sum
defined in \eref{Sig-Def} and its final expression is given in
(\ref{Sig-E}). If we put all $Q_i\to K$ and $T\to R$ in
(\ref{Sig-E}) we will get box coefficients.

However, there is a technical issue related to the $u$-dependence
(i.e., the $\mu^2$-dependence part, which indicates the dimensional
shifted basis) contained inside the definition of $R$. To have clear
separations of $u$-dependence, we will expand $R$ smartly. To do so,
 we  construct the  vector
$q_0^{q_i,q_j,K}$ orthogonal to all three momenta $K_i, K_j, K$
\bea
{(q_0)}_{\mu}^{q_i,q_j,K}={1\over K^2}\epsilon_{\mu \nu\rho\xi}
q_i^{\nu}q_j^\rho K^\xi={1\over K^2}\epsilon_{\mu \nu\rho\xi}K_i^{\nu}K_j^\rho K^\xi.
\eea
Then we can expand $R$ using $Q_i,Q_j,K,{(q_0)}_{\mu}^{q_i,q_j,K}$
and obtain
\bea
\Spab{P_1|R|P_2}=\b\sqrt{1-u}a_0^{(q_i,q_j,q_0;r)}\Spab{P_1|q_0|P_2}+\beta^{(q_i,q_j,q_t;r)}\Spab{P_1|K|P_2}
~~\label{R-exp-box-another} \eea
with
\bea a_0^{(q_i,q_j,q_0;r)}&=&{r\cdot q_0^{(q_i,q_j,K)}\over
(q_0^{(q_i,q_j,K)})^2}~~~\label{ab-zero}\\
\beta^{(q_i,q_j,q_0;r)}&=&{N^{(K_i,K_j,K;\wt R)}\over K^2
{G(K_i,K_j,K)}}~~~\label{beta-zero} \eea
where $N$ and $G$ are defined in \eref{N-4var} and
\eref{Gram-def-1}. The expression \eref{R-exp-box-another} has the clear
$u$-dependence.

Putting  the expansion (\ref{R-exp-box-another}) back,  we have
\bea {\Spab{P_1|R|P_2}^{s}\over
\Spab{P_1|K|P_2}^{s}}=\sum_{h=0}^{s}{s\choose
h}{a_0^{(q_i,q_j,q_0;r)}}^h{\beta^{(q_i,q_j,q_0;r)}}^{s-h}
{(\beta\sqrt{1-u})^h\Spab{P_1|q_0|P_2}^{h}\over
\Spab{P_1|K|P_2}^{h}}. ~~~\label{Box-D}\eea
Summing  the above result with  the term coming from  exchanging $P_1$ and
$P_2$ and using  the formula  (\ref{Sig-E}) in the Appendix, we have
\bea &&{(\beta\sqrt{1-u})^h\Spab{P_1|q_0|P_2}^{h}\over
\Spab{P_1|K|P_2}^{h}}
+{(\beta\sqrt{1-u})^h\Spab{P_2|q_0|P_1}^{h}\over
\Spab{P_2|K|P_1}^{h}}\nn &=& \left\{
  \begin{array}{ll}
    {2(2i)^h(q_0^2)^h\{\beta^2(1-u)[(2q_i\cdot q_j)^2-4q_i^2q_j^2]
    +4K^2[\alpha_i\alpha_j(2q_i\cdot q_j)-\alpha_i^2q_j^2-\alpha_j^2q_i^2]\}^{h/2}
\over [(2q_i\cdot q_j)^2-4q_i^2q_j^2]^h}, & \hbox{for $h$ even;} \\
    0, & \hbox{for $h$ odd.}
  \end{array}
\right.
\eea
Thus
\bea &&{\Spab{P_1|R|P_2}^{s}\over
\Spab{P_1|K|P_2}^{s}}+\{P_1\leftrightarrow P_2\}
=\sum_{even~h=0}^{s}{s\choose h}{ {N^{(K_i,K_j,K;\wt
R)}}^{s-h}2T\over (K^2 G(K_i,K_j,K))^s},~~~~\label{Box-Sim} \eea
where\footnote{Also $\b,u, \a_i,\a_j$ have $K^2$ in
denominators. It can be checked that the overall $T$ does not have $K^2$
in the denominator. This is important, because the box coefficient
(\ref{Box-L-Coef}) will not have $K^2$ as its singularity.}
{ \bea  T(h)&=&
\left\{\beta^2(1-u)G(K_i,K_j,K)-K^2\left[2\alpha_i\alpha_j G\left(
 \begin{array}{cc}
  K& K_j \\
     K & K_i \\
   \end{array}
   \right) -\alpha_i^2G(K,K_j) -\alpha_j^2 G(K,K_i)
                        \right]\right\}^{h/2}\nn & & \left(-K^2G(K_i,K_j,\wt R,K)\right)^{h/2}.~~~~~\label{T-def}
\eea}

Combining  Eq.(\ref{Box-ind}) and (\ref{Box-Sim})  the  box
coefficients are given by
\bea
C[Q_i,Q_j,K]&=&\sum_{z_1+\ldots+z_k+s=n+2}\sum_{h=0}^s\left(\prod_{t=1,t\neq
i,j}^k {G\left( \begin{array}{cccc} K_i & K_j & K_t & K \\
K_i & K_j & \W R & K
\end{array}\right) \over G(K_i,K_j,K_t,K)}
\left({N^{(K_i,K_j,K_t,K;\wt R)}\over
G(K_i,K_j,K_t,K)}\right)^{z_t}\right)\nn &&\times {s\choose h}{
{N^{(K_i,K_j,K;\wt R)}}^{s-h}T(h)\over (G(K_i,K_j,K))^s};~~~\hbox{with
$h$ even}~~~\label{Box-L-Coef} \eea
where $T$ is defined in Eq.({\ref{T-def}}). The analysis of
the singularity structure of \eref{Box-L-Coef} has been given in the
first subsection.

\section{Coefficients of triangle}
The triangle coefficient is given in (\ref{Tri-coeff-1}) and we
recall here that when $n\geq -1$,
\bea C[Q_i,K]&=&{(K^2)^{n+1}\over 2}{1\over
(\sqrt{\Delta})^{n+1}}{1\over (n+1)!\Spaa{P_1~P_2}^{n+1}}\nn
&&\times \left.{d^{n+1}\over d \tau^{n+1}}\left(\left.{\Spaa{\ell|R
Q_i|\ell}^{n+k}\over \prod_{t=1,t\neq i}^k\Spaa{\ell|Q_t
Q_i|\ell}}\right|_{\ell\to P_1-\tau P_2} +\{P_1\leftrightarrow
P_2\}\right)\right|_{\tau \to 0}~~~~\label{Tri-C} \eea
where $P_1,P_2$ are two massless momenta constructed from $Q_i,K$ as
given in \eref{P1P2}. More explicitly $P_1$ and $P_2$ are given by
\bea P_{1,2}=Q_i+x_{1,2}K~~~\label{C-Form-Def} \eea
with
\bea x_{1,2}={-2\alpha_i K^2 \pm\sqrt{\Delta}\over 2K^2},~~~
\sqrt{\Delta}=\beta\sqrt{1-u}\sqrt{\delta},\quad \delta=-4q_i^2K^2.
~~~\label{Tri-x12}\eea
To have a good separation of the $u$-dependence, we can also construct
two massless momenta $p_{1,2}$ from $q_i$ and $K$
\bea p_{1,2}&=&q_i+y_{1,2}K, \quad y_{1,2}=\pm {\sqrt{\delta}\over
2K^2}~~~~\label{O-Form-Def}~. \eea
Comparing definitions (\ref{C-Form-Def}) with (\ref{O-Form-Def}) we
have
\bea P_{1,2}
=\beta\sqrt{1-u}p_{1,2},~~~~\label{P-p-rel} \eea
thus the triangle coefficient (\ref{Tri-C}) \cite{Britto:2008FY} can
be rewritten as
\bea C[Q_i,K]&=&{(K^2)^{n+1}\over 2}{1\over
(\sqrt{\delta})^{n+1}}{1\over (n+1)!\Spaa{p_1~p_2}^{n+1}}\nn
&&\times \left.{d^{n+1}\over d
\tau^{n+1}}\left(\left.{\Spaa{\ell|\wt r Q_i|\ell}^{n+k} \over
\prod_{t=1,t\neq i}^k\Spaa{\ell|\wt q_t Q_i|\ell}}\right|_{\ell\to
p_1-\tau p_2} +\{p_1\leftrightarrow p_2\}\right)\right|_{\tau \to
0},~~~~\label{Tri-C-F1} \eea
where
\bea \wt r=r-{\alpha_R\over \alpha_i}q_i,\quad \wt
q_t=q_t-{\alpha_t\over \alpha_i}q_i. \eea
The good property of expression  (\ref{Tri-C-F1}) is that  only
$Q_i$s have the  $u$-dependence.  Now we will evaluate residues
based on this expression  (\ref{Tri-C-F1}).

The presentation of this section is following. In the first
subsection we present the result and analyze the singularity
structure. For readers who cares only the result, reading this subsection
is enough. In subsection 5.2, we will evaluate the derivative part
in  the expression \eref{Tri-C-F1} and finally we give  the Lorentz-invariant
form in subsection 5.3, where the polynomial property of $u$ is a natural by-product.

\subsection{The summary of main results of current section }

The Lorentz invariant form of external momenta of the triangle
coefficient is given by
\bea C[Q_i,K]&=&{(K^2)^{2(n+1)}\over
(-2)^{n+1}}\sum_{s=0}^{n+1}\sum_{s'=0}^{[s/2]}
\sum_{\stackrel{\{z_1,z_2,...,z_k\}\geq 0}{  \sum_{t=1,t\neq i}^k
z_t=n+1-s}} {(n+k)!T_1(s,s')T_2(z_t)\over s'!(s-2s')!(n+k-s+s')!}\nn
&&\times \left(\prod_{t=1,t\neq
i}^k{1\over{G(K_i,K_t,K)}^{1+z_t}}\right)\left(
\sum_{\stackrel{\{h_1,h_2,...,h_k\}\geq 0}{  \sum_{t=1,t\neq i}^k
h_t={\rm even}}}^{\{ 1+z_1,1+z_2,...,1+z_k\}}
{T_3(z_t,h_t)\over {G(K_i,K)}^{n+1-h/2}}\right),
~~~\label{Tri-Loren-F-1} \eea
where $h=\sum_{t=1,t\neq i}^k h_t$. Various functions $T_1, T_2,
T_3$ can be found in \eref{T-L-Def} and $G$ is the Gram determinant
defined in \eref{Gram-def} and \eref{Gram-def-1}.

From (\ref{Tri-Loren-F-1}) we can easily read out the analytic
structure of triangle coefficients:
\begin{itemize}

\item First the coefficient contains only second-type singularities
and the first-type singularity related to the triangle topology appears
only in the triangle scalar basis (with dimensional shifted basis).

\item There are only two kinds of second-type singularities. The
first kind of second-type singularities is given by $G(K_i,K)=0$,
which is the second-type singularity intrinsically related to the
triangle topology specified by momenta $K_i, K$. The highest degree
of the pole $G(K_i,K)$ is $(n+1)$. It fits with the naive observation in the
reduction, i.e., among $(n-k)$ inner momenta $\W\ell$ in the numerator,
$(k-1)$ of them have been used to remove  $(k-1)$ propagators,
thus it is left with $(n+1)$ $\W\ell$ in numerator contributing to
the triangle topology. Each $\W\ell$ will bring one factor
$G(K_i,K)$ in the denominator of the coefficient, thus we will have the
degree $(n+1)$.

\item For the pole $G(K_i,K,K_t)$, which
is the second-type singularity intrinsically related to the box
topology specified by momenta $K_i,K_t, K$, the highest degree is
$(n+2)$. Its appearance is very natural since these boxes can be
reduced to triangled by pinching one propagator. In other words,
their influence to the triangle is given by the appearance of the factor
$G(K_i,K_t,K)$. Moreover, it fits with the naive observation in the
reduction and is, in fact, the same highest degree found for the box
coefficient in the previous section. The same highest degree $(n+2)$ is
also necessary for the cancelation of soft or collinear singularities
between box and triangle contributions.

It is worth to mention that when $k=1$, there is no box coefficient
at all. From (\ref{Tri-Loren-F-1}), we can see that the pole
$G(K_i,K,K_t)$ will not appear, which is consistent.

\item Similarly to the box case, we need to check the
$u$-dependence part in the numerator. From \eref{Tri-Loren-F-1}, all
$u$-dependence comes from factors $T_1, T_2$ and its highest degree
is $[(n+1)/2]$. It is also important to notice that each $u$
will be accompanied by a factor $G(K_i,K)$ (see
Eq.({\ref{T-L-Def}})), which will reduce the degree of the pole
$G(K_i,K)$ for these (rational) parts.

\end{itemize}
%

\subsection{Evaluation of the derivative part}
%
The \eref{Tri-C-F1} contains the standard sum defined in
\eref{Sig-Def}, but there is also the differential action. Thus to apply
the result in Appendix, we need to evaluate the derivative part first.
Let us define
\bea
f=\Spaa{\ell|\wt r Q_i|\ell}^{n+k},\quad g={1\over \prod_{t=1,t\neq i}^k\Spaa{\ell|\wt q_t Q_i|\ell}},
\eea
then the derivative gives
\bea
 {d^{n+1}(fg)\over d \tau^{n+1}}=\sum_{s=0}^{n+1}{n+1\choose
 s}f^{(s)}g^{(n+1-s)},
\eea
where $(*)^{(s)}$ denote the  $s$-th order derivative of the
function $(*)$. The $Q_i$ is a linear combination of $p_{1,2}$
 \bea Q_i
&=&\mu_1p_1+\mu_2 p_2,~~~~\mu_{1,2}={\beta\sqrt{1-u}\over
2}\pm{\alpha_i\over 2y_1}~~~\label{mu-12} \eea

~\\

{\bf The evaluation of $f^{(s)}$}:
After some algebraic manipulations, we can easily get
\bea
\Spaa{p_1-\tau p_2|\wt r Q_i|p_1-\tau p_2}
=\Spaa{p_1~p_2}a_0(\tau-\tau_{0,1})(\tau-\tau_{0,2})
\eea
where
\bea a_0&=&\mu_1 \Spab{p_2|\wt r|p_1},\quad
\tau_{0,1}={{\alpha_i\over y_1}(2\wt r\cdot q_i)+\sqrt{\Omega(\wt
r)}\over 2a_0} \quad \tau_{0,2}={{\alpha_i\over y_1}(2\wt r\cdot
q_i)-\sqrt{\Omega(\wt r)}\over 2a_0}.~~~\label{a-tau} \eea
and
\bea \Omega(\wt r)&=&(\mu_2\Spab{p_2|\wt r|p_2}-\mu_1 \Spab{p_1|\wt
r|p_1})^2 +4\mu_1\mu_2 \Spab{p_2|\wt r|p_1} \Spab{p_1|\wt r|p_2}\nn
&=&{\alpha_i^2\over y_1^2}(2\wt r\cdot q_i)^2+4\left(\beta^2(1-u)-{\alpha_i^2\over y_1^2}\right)((q_i\cdot \wt r)^2-q_i^2\wt r^2)
\eea
with the explicit $u$-dependence.
%
%

To continue, we need the following formula
\bea (b_1b_2\dots b_n)^{(k)}=\sum_{z_1+z_2+\dots+z_n=k}{k!\over
z_1!z_2! \dots z_n!}b_1^{(z_1)}b_2^{(z_2)}\dots
b_n^{(z_n)}.~~~\label{Poly-der} \eea
After  we set $b_1=b_2=\dots=a_0(\tau-\tau_{0,1})(\tau-\tau_{0,2})$,
to have nonzero result,  $0\leq z_j\leq 2$. Using $s'$ to denote the
number of $b_i$ having  the second-order derivative (so there must be
$(k-2s')$ of $b_i$ having the first-order derivative), we have
\bea
(b_1b_2\dots b_n)^{(k)}
&=&\sum_{s'=0}^{[k/2]}{n \choose s'}{n-s' \choose k-2s'}k!(a_0)^{s'}
[a_0(-\tau_{0,1}-\tau_{0,2})]^{k-2s'}[a_0\tau_{0,1}\tau_{0,2})]^{n-k+s'}, \eea
where we have take $\tau \to 0$.
Substituting  $n\to n+k, k\to s$ into the above result, we obtain
\bea f^{(s)}
&=&\Spaa{p_1~p_2}^{n+k}\sum_{s'=0}^{[s/2]}{n+k\choose
s'}{n+k-s'\choose s-2s'}s!
\left(-\left(\beta^2(1-u)-{\alpha_i^2\over y_1^2}\right){(\wt
r|\wt r)\over K^2 }\right)^{s'}\nn &&\times \left(-{\alpha_i\over
y_1}(2\wt r\cdot q_i)\right)^{s-2s'}(-\mu_2
 \Spab{p_1|\wt r|p_2})^{n+k-s}~~\label{T-f}
\eea
where we have used the short notation $(\wt r_1|\wt r_2)\equiv(\wt
r_1|\wt r_2)_{q_i,K}$ defined in \eref{Gram-def-2} since in this
section, $q_i,K$ are fixed.

~\\

{\bf The evaluation of $g^{(n+1-s)}$}:
Similar to $f$, $g$ can be written as
\bea
g
&=&{1\over \Spaa{p_1~p_2}^{k-1}}\prod_{t=1,t\neq i}^k {1\over a_t(\tau_{t,1}-\tau_{t,2})}\left({1\over \tau-\tau_{t,1}}-{1\over \tau-\tau_{t,2}}\right)
\eea
where $a_t$, $\tau_{t,1}$ and$\tau_{t,2}$ are given in (\ref{a-tau})
 with $\wt
r$ replaced by $\wt q_t$. Then from Eq.({\ref{Poly-der}}) we can get
\bea g^{(n+1-s)}&=&{(n+1-s)!\over
\Spaa{p_1~p_2}^{k-1}}\sum_{\stackrel{\sum_{t=1,t\neq i}^k
z_t=n+1-s}{z_t\geq 0}} \prod_{t=1,t\neq i}^k{1\over
z_t!a_t(\tau_{t,1}-\tau_{t,2})} \left({1\over
\tau-\tau_{t,1}}-{1\over \tau-\tau_{t,2}}\right)^{(z_t)}\nn
%
%
&=&{(n+1-s)!\over
\Spaa{p_1~p_2}^{k-1}}\sum_{\stackrel{\sum_{t=1,t\neq i}^k
z_t=n+1-s}{z_t\geq 0}} \prod_{t=1,t\neq i}^k{1\over
\sqrt{\Omega(q_t)}} \left({1\over \tau_{t,2}^{1+z_t}}-{1\over
\tau_{t,1}^{1+z_t}}\right),\quad \tau\to 0. \eea
%
%
Substituting expressions of $\tau_{t,1}$ and $\tau_{t,2}$ yields
\bea \label{T-g} g^{(n+1-s)}&=&{(n+1-s)!\over
\Spaa{p_1~p_2}^{k-1}}\sum_{\stackrel{\sum_{t=1,t\neq i}^k z_t=n+1-s}
{z_t\geq 0}} \prod_{t=1,t\neq i}^k
{\sum_{\gamma_t=0}^{[z_t/2]}2{1+z_t\choose 2\gamma_t+1}( \Omega(\wt
q_t))^{\gamma_t}\left({\alpha_i\over y_1}(2\wt q_t\cdot
q_i)\right)^{z_t-2\gamma_t} \over (-2\mu_2\Spab{p_1|\wt
q_t|p_2})^{1+z_t}} \eea
%

%

~\\
{\bf The final result of derivation:}
Putting all together   and performing a bit  algebraic
manipulations, we can write Eq.(\ref{Tri-C-F1}) as
\bea C[Q_i,K]
&=&{(K^2)^{n+1}\over 2}{1\over
(-2q_i^2)^{n+1}}\sum_{s=0}^{n+1}\sum_{s'=0}^{[s/2]}
\sum_{\stackrel{\sum_{t=1,t\neq i}^k z_t=n+1-s}{z_t\geq  0}}
{(n+k)!\over s'!(s-2s')!(n+k-s+s')!}\nn &&\times
T_1(s,s')T_2(z_t)\left({\Spab{p_1|\wt r|p_2}^{n+k-s}\over
\prod_{t=1,t\neq i}^k\Spab{p_1|\wt q_t|p_2}^{1+z_t}
}+\{p_1\leftrightarrow p_2\}\right), ~~~\label{Tri-Spin-F} \eea
where the Lorentz invariant forms of $T_1, T_2$ are
\bea T_1(s,s')&=&\left(\left(\alpha_i^2-y_1^2\beta^2(1-u)\right)
{(\wt r|\wt r)\over K^2 }\right)^{s'}(-\alpha_i(2\wt r\cdot
q_i))^{s-2s'},\nn T_2(z_t)&=& \prod_{t=1,t\neq i}^k
\sum_{\gamma_t=0}^{[z_t/2]}{1+z_t\choose 2\gamma_t+1}( {1\over
4}y_1^2\Omega(\wt q_t))^{\gamma_t} (\alpha_i(\wt q_t\cdot
q_i))^{z_t-2\gamma_t}.~~~\label{T-N-Def} \eea
The $u$-dependence is entirely in $T_1$ and $T_2$, thus the
polynomial property of $u$ is obvious.
\subsection{The Lorentz-invariant Form}

In \eref{Tri-Spin-F},  the sum inside the bracket
is the standard one defined in Appendix \eref{Sig-Def}. Thus we can
use the result (\ref{Sig-E}) given in  the Appendix. First noticing
that $\Spab{p_1| q_i|p_2}=0$ and $\Spab{p_2| q_i|p_1}=0$ by our
construction \eref{O-Form-Def}, $\Spab{p_1|\wt q_t|p_2}=\Spab{p_1|
q_t|p_2}$ and $\Spab{p_1|\wt r|p_2}=\Spab{p_1| r|p_2}$ in
\eref{Tri-Spin-F}. After some algebraic calculations,
Eq.(\ref{Tri-Spin-F}) leads to
\bea C[Q_i,K]&=&{(K^2)^{2(n+1)}\over
(-2)^{n+1}}\sum_{s=0}^{n+1}\sum_{s'=0}^{[s/2]}
\sum_{\stackrel{\{z_1,z_2,...,z_k\}\geq 0}{  \sum_{t=1,t\neq i}^k
z_t=n+1-s}} {(n+k)!T_1(s,s')T_2(z_t)\over s'!(s-2s')!(n+k-s+s')!}\nn
&&\times \left(\prod_{t=1,t\neq
i}^k{1\over{G(K_i,K_t,K)}^{1+z_t}}\right)\left(
\sum_{\stackrel{\{h_1,h_2,...,h_k\}\geq 0}{  \sum_{t=1,t\neq i}^k
h_t={\rm even}}}^{\{ 1+z_1,1+z_2,...,1+z_k\}} {
T_3(z_t,h_t)\over {G(K_i,K)}^{n+1-h/2}}\right),
~~~\label{Tri-Loren-F} \eea
where $h=\sum_{t=1,t\neq i}^k h_t$, $G$ is Gram determinant defined
in \eref{Gram-def} and \eref{Gram-def-1} and
\bea
T_1(s,s')&=&\left(\Omega_1(\wt R)\right)^{s'}
\left(2\Omega_2(\wt R)\right)^{s-2s'},\nn
T_2(z_t)&=& \prod_{t=1,t\neq i}^k
\sum_{\gamma_t=0}^{[z_t/2]}{1+z_t\choose 2\gamma_t+1} \left(
\Omega_1(K_t)+{(\Omega_2(
K_t))}^2\right)^{\gamma_t}(-\Omega_2(K_t))^{z_t-2\gamma_t},\nn
T_3(z_t,h_t)&=&\prod_{t=1,t\neq i}^k{1+z_t\choose h_t}{\left(\epsilon(\wt
R,K_i,K_t,K)\right)}^{h_t}
{\left(G\left( \begin{array}{ccc} K_i & K & K_t  \\
K_i & K & \W R
\end{array}\right)\right)}^{1+z_t-h_t}~~~\label{T-L-Def}, \eea
with %
\bea \Omega_1(\wt R)&=&\left(\alpha_i^2+{G(K_i,K)\over
(K^2)^2}\beta^2(1-u)\right) {G(K_i,\wt R,K)\over K^2 },\nn
\Omega_2(\wt R)&=&{1\over K^2} \left( \a_R G(K,K_i)-\a_i
                         G\left( \begin{array}{cc} K & K_i   \\
K  & \W R
\end{array}\right)\right).
\eea
It is worth to mention that it can be checked that $\Omega_1$ has
$(K^2)^4$ in the denominator and $\Omega_2$ has $(K^2)^2$ in the
denominator. When putting back into \eref{Tri-Loren-F}, the $K^2$ factor
from $\Omega_1, \Omega_2$ will be canceled by the overall
$(K^2)^{2(n+1)}$ factor. In other words, $K^2$ will not be a
singularity for the triangle coefficient.

Since our main concern is the highest degrees of poles $G(K_i,K)$ and
$G(K_i,K_t, K)$, we will discuss how to get this information from
\eref{Tri-Loren-F}. For the pole $G(K_i,K_t, K)$ there is one and only
one term with highest degree in the expression \eref{Tri-Loren-F}
which is given by $s=0, s'=0$, $z_t=n-1$ and $z_r=0$ for $r\neq
i,t$. In other words, the highest degree of the pole $G(K_i,K_t, K)$ is
$(n+2)$.

For the pole $G(K_i,K)$, since there are many terms contributing to
the highest degree in the expression \eref{Tri-Loren-F}, it will be a
little more complicated and we will use another expression to
discuss. Since $G(K_i,K)=K^2 q_i^2$, we will rewrite
\eref{Tri-Loren-F} using $q_i^2$. Noting that
\bean T_1(s,s')&=&2^{s-2s'}(-\alpha_i(r\cdot q_i))^{s},\nn
 T_2(z_t)&=& \prod_{t=1,t\neq i}^k
({1+z_t})(\alpha_i(q_t\cdot q_i))^{z_t}.\nn
 \eean
therefore, after removing terms with $q_i^2$ in the numerator, we
have
\bea C[Q_i,K] &\to &{(K^2)^{2(n+1)}\alpha_i^{n+1}(-(r\cdot
q_i))^{n+k}\over
(-2)^{n+1}(q_i^2)^{n+1}}\sum_{s=0}^{n+1}\sum_{s'=0}^{[s/2]}
{(n+k)!2^{s-2s'}\over s'!(s-2s')!(n+k-s+s')!}\nn &&\times
\sum_{\stackrel{\{z_1,z_2,...,z_k\}\geq 0}{  \sum_{t=1,t\neq i}^k
z_t=n+1-s}} \left(\prod_{t=1,t\neq i}^k\left({({1+z_t})(q_t\cdot
q_i)^{1+2z_t}\over {(q_i^2q_t^2-(q_i\cdot
q_t)^2)^{1+z_t}}}\right)\right)~~~~\label{Tri-Loren-F-2}
 \eea
With this explicit form, the highest degree of the pole $q_i^2$ (or the pole
$G(K_i,K)$) is $(n+1)$.

\section{Coefficients of bubble}
After accomplishing the triangle coefficients, the last thing is to
derive the  coefficient of the  bubble. The bubble coefficient is the sum of the
residues of the poles from the following expression
\bea B&=&\sum_{i=1}^k\sum_{q=0}^{n}{-(K^2)^{n+1}\Spaa{\ell|R
Q_i|\ell}^{n-q+k-1}\over
\Spaa{\ell|KQ_i|\ell}^{n-q+1}\prod_{t=1,t\neq
i}^k\Spaa{\ell|Q_tQ_i|\ell}} {1\over
q+1}{\Spab{\ell|R|\ell}^{q+1}\over
\Spab{\ell|K|\ell}^{q+1}}.~~\label{Coeff-Bubble-1} \eea
This expression is for $k\geq 1$. For $k=0$, the answer is very
simple and we write down here\footnote{With the definition of \eref{r-q-def} and
\eref{ab-def}, we can see the overall $(K^2)^{-n}$- dependence, which is the only singularity
for the case $k=0$.}
\bea C[K]_{k=0} & = & {\sum_{z=0}^{[(n+1)/2]} (-2\a_R K^2)^{n-2z}
(-4\b^2 (1-u) G(K,\W R))^z\over 2^{n+1}
(n+1)}~~\label{Coeff-Bubble-k0}\eea

As in previous two sections, we summarize the final result and discuss the analytic
property in the first subsection. The  derivation of the result is given in the next three
subsections. In subsection 6.2, we present the explicit spinor form after the evaluations of
residues. In subsection 6.3, we deal with the derivative part and finally in subsection 6.4, we
translate the spinor form into the Lorentz-invariant form.

\subsection{The summary of main result of current section}

The Lorentz invariant form of bubble coefficient is given by
\bea C[K]&=& \sum_{i=1}^k\sum_{q=0}^{n}{(-1)^{n-q}(K^2)^{2n+1-q}}
\sum_{s=
Max\{n-2q-1,0\}}^{n-q}\sum_{s_1=0}^{s}\sum_{s'_1=0}^{[s_1/2]}
T_0(s,s_1,s_1')T_1(s_1,s_1')\nn & & \left(-{{G(K_i,\wt R,K)}\over K^2
}\right)^{n-q-s}\times \sum_{\stackrel{\sum_{t=1,t\neq i}^k
z_t=s-s_1}{z_t\geq 0}}
\sum_{r_1=0}^{2q+1+s-n}{\sum_{r_2=0}^{n-q-s}}{T_2(z_t)T_4(r_1,r_2)\over
(-4{G(K_i,K)})^{(n-q-s+1+r_1-r_2)/2}}\nn
 &&\times\left(2\prod_{t=1,t\neq
i}^k{1\over{G(K_i,K_t,K)}^{1+z_t}}\right)\left(
\sum_{\stackrel{\{h_1,h_2,...,h_k\}\geq 0}{  \sum_{t=1,t\neq i}^k
h_t={\rm even}}}^{\{ 1+z_1,1+z_2,...,1+z_k\}} {
T_3(z_t,h_t)\over {G(K_i,K)}^{n-q-h/2}}\right)~~~\label{Bubb-Coeff-1} \eea
where $T_0, T_4$ can be found in \eref{T0T4} and $T_1, T_2,T_3 $ are
defined in Eq.(\ref{T-N-Def}). In the sum \eref{Bubb-Coeff-1}, we need
to have $(n-q-s+1+r_1-r_2)$ and $h=\sum_{t=1,t\neq i}^k h_t$ to be even number.

From \eref{Bubb-Coeff-1}, the analytic property of bubble coefficients
can be read out as follows\footnote{The derivation of the highest
degree can be found in  subsection 6.4.}:
\begin{itemize}

\item Like  coefficients of the box and triangle, only second-type
singularities appear in bubble coefficients (remembering the first-type
singularity of the bubble is $(K^2-(M_1\pm M_2)^2)$). There are three
second-type singularities:  $G(K,K_i,K_j)$ related to the box
topology,  $G(K,K_i)$ related to the triangle topology and $K^2$
related to the bubble topology.

\item The highest degree of the pole $K^2$ is $n$. $K^2$ is the intrinsic
second-type singularity related to the  bubble topology. Its highest
degree fits the naive observation in the reduction: to remove $k$
propagators from the denominator we need to reduce the $k$'s $\W \ell$
in the numerator.

\item The highest degree of the pole $G(K,K_i)$ is $(n+1)$, which fits
with the naive observation in the reduction. It is also the same as
 the pole $G(K,K_i)$ appearing in triangle coefficients. Having
the same highest degree is reasonable when we consider some soft or collinear
limits of full one-loop amplitudes after combining all contributions (such as box, triangle and bubble)
 together.

\item The highest degree of the pole $G(K,K_i,K_j)$ is $(n+1)$. It is
worth to recall that the highest degree of the same pole in box and
triangle coefficients is $(n+2)$. This is because to
get the influence of the box to the bubble, one further reduction from
the triangle to the bubble  is needed.

\item To see the dimensional shifted basis (which is related to rational
part of one-loop amplitudes \eref{D-basis}), we need to check the
$u$-dependence part in the numerator. The highest degree of $u$ is ${[n/2]}$.

\end{itemize}
%

\subsection{Spinor form of the bubble coefficient}
The expression \eref{Coeff-Bubble-1} is not yet the spinor form of
the coefficient of the bubble. We need to evaluate residues of various poles
from $\Spaa{\ell|KQ_i|\ell}$ and $\Spaa{\ell|Q_jQ_i|\ell}$. Among these
two kinds of poles, the contributions of poles
from $\Spaa{\ell|Q_jQ_i|\ell}$ are zero. To see it, let us start with the
typical term
\bea
B_{i,q}&\equiv&{-(K^2)^{n+1}\Spaa{\ell|R Q_i|\ell}^{n-q+k-1}\over
\Spaa{\ell|KQ_i|\ell}^{n-q+1}\prod_{t=1,t\neq i}^k\Spaa{\ell|Q_tQ_i|\ell}}
{1\over q+1}{\Spab{\ell|R|\ell}^{q+1}\over \Spab{\ell|K|\ell}^{q+1}}.
\eea
and construct two massless momenta as (see (\ref{Pole-construct}))
\bea
P_{1,2}^{(i,j)}&=&Q_j+y_{1,2}^{(i,j)}Q_i,\quad (i<j)
\eea
where
\bea
y_{1,2}^{(i,j)}={-2Q_i\cdot Q_j\pm \sqrt{\Delta^{(i,j)}}\over 2Q_i^2},\quad \Delta^{(i,j)}=(2Q_i\cdot Q_j)^2-4Q_i^2Q_j^2.
\eea
Then the  residues of the poles $P_{1,2}^{(i,j)}$ are
{\small\bean {\rm Res}(B_{i,q})|_{P_{1,2}^{(i,j)}}&=&\pm{1\over
\sqrt{\Delta^{(i,j)}}}{(K^2)^{n+1}\Spab{P_{1,2}^{(i,j)}|R|P_{2,1}^{(i,j)}}^{n-q+k-1}\over
 \Spaa{P_{1,2}^{(i,j)}|K|P_{2,1}^{(i,j)}}^{n-q+1}\prod_{t=1,t\neq i,j}^k\Spaa{P_{1,2}^{(i,j)}|Q_t|P_{2,1}^{(i,j)}}}
{1\over q+1}{\Spab{P_{1,2}^{(i,j)}|R|P_{1,2}^{(i,j)}}^{q+1}\over
\Spab{P_{1,2}^{(i,j)}|K|P_{1,2}^{(i,j)}}^{q+1}} \eean}
where $(+)$-sign is  for the pole $P_{1}^{(i,j)}$ and $(-)$-sign, for
the pole $P_{1}^{(i,j)}$. It is worth to notice that the factor
$\Spaa{\ell|Q_jQ_i|\ell}$ appears in both $B_{i,q}$ and $B_{j,q}$ up
to a minus sign, thus ${\rm Res}(B_{i,q})|_{P_{1,2}^{(i,j)}}=-{\rm
Res}(B_{j,q})|_{P_{1,2}^{(i,j)}}$. So when we sum up all residues,
contributions from $\Spaa{\ell|Q_jQ_i|\ell}$ cancel.

Now we consider  poles from $\Spaa{\ell|KQ_i|\ell}$, which has been
carefully discussed  in Eq.({\ref{C-Form-Def}}). The residue of
$P_1$ is given by
\bean {\rm
Res}(B_{i,q})|_{P_1}&=&\left.{(-1)^{n-q}(K^2)^{2n+2-q}(x_1-x_2)^{n-q+1}\over
\Spaa{P_1~P_2}^{n-q}\Delta^{n-q+1}(q+1)(n-q)!}{d^{n-q}\over
d\tau^{n-q}} \left({\Spaa{\ell|R
Q_i|\ell}^{n-q+k-1}\over\prod_{t=1,t\neq i}^k\Spaa{\ell|Q_t
Q_i|\ell}}{\Spab{\ell|R|P_1}^{q+1}\over
\Spab{\ell|K|P_1}^{q+1}}\right)\right|_{\ell \to P_1-\tau P_2}, \eean
and the residue of $P_2$
\bean {\rm
Res}(B_{i,q})|_{P_2}&=&\left.{(-1)^{n-q+1}(K^2)^{2n+2-q}(x_1-x_2)^{n-q+1}\over
\Spaa{P_1~P_2}^{n-q}\Delta^{n-q+1}(q+1)(n-q)!}{d^{n-q}\over
d\tau^{n-q}} \left({\Spaa{\ell|R
Q_i|\ell}^{n-q+k-1}\over\prod_{t=1,t\neq i}^k\Spaa{\ell|Q_t
Q_i|\ell}}{\Spab{\ell|R|P_2}^{q+1}\over
\Spab{\ell|K|P_2}^{q+1}}\right)\right|_{\ell \to P_2-\tau P_1}. \eean
Using the relation (\ref{P-p-rel}) and the corresponding $p_{1,2}$
in Eq.({\ref{O-Form-Def}}),  similarly to the case of the triangle, the
above two equations can be simplified as:
\bea {\rm Res}(B_{i,q})|_{P_1}&=&{(-1)^{n-q}(K^2)^{n+1}\over
\beta(\sqrt{1-u})\Spaa{p_1~p_2}^{n-q}\sqrt{\delta}^{n-q+1}(q+1)(n-q)!}\nn
&&\times {d^{n-q}\over d\tau^{n-q}}\left.\left({\Spaa{\ell|\wt r
Q_i|\ell}^{n-q+k-1}\over\prod_{t=1,t\neq i}^k\Spaa{\ell|\wt q_t
Q_i|\ell}} {\Spab{\ell|\beta(\sqrt{1-u})r+\alpha_R K|p_1}^{q+1}
\over \Spab{\ell|K|p_1}^{q+1}}\right)\right|_{\ell \to p_1-\tau
p_2}, \eea
and,
\bea {\rm Res}(B_{i,q})|_{P_2}&=&{(-1)^{n-q+1}(K^2)^{n+1}\over
\beta(\sqrt{1-u})\Spaa{p_1~p_2}^{n-q}\sqrt{\delta}^{n-q+1}(q+1)(n-q)!}\nn
&&\times {d^{n-q}\over d\tau^{n-q}}\left.\left({\Spaa{\ell|\wt r
Q_i|\ell}^{n-q+k-1}\over\prod_{t=1,t\neq i}^k\Spaa{\ell|\wt q_t
Q_i|\ell}} {\Spab{\ell|\beta(\sqrt{1-u})r+\alpha_R K|p_2}^{q+1}
\over \Spab{\ell|K|p_2}^{q+1}}\right)\right|_{\ell \to p_2-\tau
p_1}. \eea
Summing all together  we have the coefficient of the bubble in the
spinor form
\bea C[K]=\sum_{i=1}^k\sum_{q=0}^{n}({\rm Res}(B_{i,q})|_{P_1}+{\rm
Res}(B_{i,q})|_{P_2}).~~~\label{Ck-form-1} \eea
%


\subsection{Evaluation of the derivative part }
%
The spinor form \eref{Ck-form-1} is complicated and we need to evaluate
the derivation first. To do so, we define
\bea
f&\equiv&\Spaa{\ell|\wt r Q_i|\ell}^{n-q+k-1},~~~ g\equiv
{1\over\prod_{t=1,t\neq i}^k\Spaa{\ell|\wt q_t Q_i|\ell}},~~~
w\equiv{\Spab{\ell|\beta(\sqrt{1-u})r+\alpha_R K|\ell}^{q+1}\over
\Spab{\ell|K|\ell}^{q+1}}. \eea
and the derivative is given by
\bea
{d^{n-q}\over d\tau^{n-q}}(f g w)
&=&\sum_{s=0}^{n-q}\sum_{s_1=0}^{s}{n-q\choose s}{s\choose s_1}f^{(s_1)}g^{(s-s_1)}w^{(n-q-s)}
\eea
%

{\bf The evaluation of $f^{(s_1)}$ and $g^{(s-s_1)}$}:
To get $f^{(s_1)}$ and  $g^{(s-s_1)}$, we can use  the result in
Subsection 4.2. If substituting $s_1, n-q+k-1$ for $s, n+k$ in Eq.
({\ref{T-f}}) we get ($\mu_{1,2}$ is given in (\ref{mu-12}))
\bea
f_{P_1}^{(s_1)}&=&\Spaa{p_1~p_2}^{n-q+k-1}\sum_{s'_1=0}^{[s_1/2]}{n-q+k-1\choose
s'_1}{n-q+k-1-s'_1\choose s_1-2s'_1}s_1!\nn &&\times \left(-{1\over
y_1}\right)^{s_1}T_1(s_1,s'_1)(-\mu_2 \Spab{p_1|\wt
r|p_2})^{n-q+k-1-s_1}, \nn
f_{P_2}^{(s_1)}&=&\left.(-1)^{s_1}f_{P_1}^{(s_1)}\right
|_{\mu_1\leftrightarrow \mu_2,p_1\leftrightarrow p_2}.~~~~~~~~~~
\eea
If substituting $s-s_1$ for $n+1-s$ in Eq. ({\ref{T-g}}) we get
\bea g_{P_1}^{(s-s_1)}&=&{(s-s_1)!\over
\Spaa{p_1~p_2}^{k-1}}\sum_{\stackrel{\sum_{t=1,t\neq i}^k
z_t=s-s_1}{0\leq z_t\leq s-s_1}}
 {T_2(z_t)\over y_1^{s-s_1}\prod_{t=1,t\neq i}^k(-\mu_2\Spab{p_1|\wt q_t|p_2})^{1+z_t}},
\nn
g_{P_2}^{(s-s_1)}&=&\left.(-1)^{s_1-s}g_{P_1}^{(s-s_1)}\right|_{\mu_1\leftrightarrow
\mu_2,p_1\leftrightarrow p_2},~ \eea
where $T_1(s_1,s'_1)$ and $T_2(z_t)$ are defined by
Eq.(\ref{T-N-Def}) and Eq.(\ref{T-L-Def}).

{\bf The evaluation of $w^{(n-q-s)}$}:
For ${\rm Res} B_{i,q}|_{P_1}$, $w$ is
\bea w_{P_1}
&=&{1\over (\sqrt{\delta})^{q+1}}(\beta(\sqrt{1-u})2r\cdot q_i-\tau \beta(\sqrt{1-u})\Spab{p_2|r|p_1}+\alpha_R \sqrt{\delta})^{q+1}.
\eea
So when $s\geq Max\{n-2q-1,0\}$
\bea w_{P_1}^{(n-q-s)}&=&{(q+1)!\over
(2q+1+s-n)!}{(-\beta\sqrt{1-u})^{n-q-s}\over
(\sqrt{\delta})^{q+1}}\nn &&\times(\beta(\sqrt{1-u})2r\cdot q_i-\tau
\beta(\sqrt{1-u}) \Spab{p_2|r|p_1}+\alpha_R
\sqrt{\delta})^{2q+1+s-n}\Spab{p_2|r|p_1}^{n-q-s},~~\label{h-B1}
\eea
and when $s< Max\{n-2q-1,0\}$, $w_{P_1}^{(n-q-s)}=0$. After setting
$\tau \to 0$, Eq. (\ref{h-B1}) becomes
\bea w_{P_1}^{(n-q-s)}&=&{(q+1)!\over
(2q+1+s-n)!}{(-\beta\sqrt{1-u})^{n-q-s}\over
(\sqrt{\delta})^{q+1}}\nn &&\times (\beta(\sqrt{1-u})2r\cdot
q_i+\alpha_R\sqrt{\delta})^{2q+1+s-n}\Spab{p_2|r|p_1}^{n-q-s},~~\label{h-1}
\eea
Similarly for  ${\rm Res} B_{i,q}|_{P_2}$, we have
\bea w_{P_2}^{(n-q-s)}&=&{(q+1)!\over
(2q+1+s-n)!}{(-\beta\sqrt{1-u})^{n-q-s}\over
(-\sqrt{\delta})^{q+1}}\nn &&\times (\beta(\sqrt{1-u})2r\cdot
q_i-\alpha_R
\sqrt{\delta})^{2q+1+s-n}\Spab{p_1|r|p_2}^{n-q-s}.~~\label{h-2} \eea
%

{\bf The final result}:
%
Now we want to sum up residues of $P_1, P_2$ of $B_{i,q}$. Up to
a common factor, their sum is given by
%
\bea
&&(\beta(\sqrt{1-u})2r\cdot q_i+\alpha_R \sqrt{\delta})^{2q+1+s-n}\mu_2^{n-q-s}
 {\Spab{p_1|\wt r|p_2}^{s+k-1-s_1}\over \prod_{t=1,t\neq i}^k (\Spab{p_1|\wt q_t|p_2})^{1+z_t}}\nn
 &&+(-1)^{n+s}(\beta(\sqrt{1-u})2r\cdot q_i-\alpha_R \sqrt{\delta})^{2q+1+s-n}\mu_1^{n-q-s}
 {\Spab{p_2|\wt r|p_1}^{s+k-1-s_1}\over \prod_{t=1,t\neq i}^k (\Spab{p_2|\wt q_t|p_1})^{1+z_t}}.
 ~~~\label{Sum-B}
\eea
Using  the binomial expansion
\bea (\beta(\sqrt{1-u})2r\cdot q_i+\alpha_R
\sqrt{\delta})^{2q+1+s-n} \left({\beta(\sqrt{1-u})\over
2}-{\alpha_i\over
2y_1}\right)^{n-q-s}=\sum_{r_1=0}^{2q+1+s-n}\sum_{r_2=0}^{n-q-s}
C_{r_1,r_2}(P_1) \eea
where
\bea C_{r_1,r_2}(P_1)&=&{2q+1+s-n \choose r_1}{n-q-s\choose
r_2}(\beta(\sqrt{1-u})2r\cdot q_i)^{r_1}(\alpha_R
\sqrt{\delta})^{2q+1+s-n-r_1}\nn
&&\times\left({\beta(\sqrt{1-u})\over
2}\right)^{r_2}\left(-{\alpha_i\over 2y_1}\right)^{n-q-s-r_2} \eea
and similar expression for
\bea C_{r_1,r_2}(P_2)&=&(-1)^{n-q-s-1+r_1+r_2}C_{r_1,r_2}(P_1), \eea
 Eq.({\ref{Sum-B}}) becomes
\bea \sum_{r_1=0}^{2q+1+s-n}\sum_{r_2=0}^{n-q-s}
C_{r_1,r_2}(p_1)\left( {\Spab{p_1|\wt r|p_2}^{s+k-1-s_1}\over
\prod_{t=1,t\neq i}^k (\Spab{p_1|\wt q_t|p_2})^{1+z_t}}
+{(-1)^{n-q-s-1+r_1+r_2}\Spab{p_2|\wt r|p_1}^{s+k-1-s_1}\over
\prod_{t=1,t\neq i}^k (\Spab{p_2|\wt q_t|p_1})^{1+z_t}}\right). \eea
Since the bubble coefficient is the polynomial of $u$,  from the
factor $ (\beta\sqrt{1-u})^{n-q-s-1+r_1+r_2}$ in the sum, only the
terms with $(n-q-s-1+r_1+r_2)$ being even numbers are left. In other
words, the above expression can be written as
\bea \sum_{r_1=0}^{2q+1+s-n}\sum_{r_2=0}^{n-q-s}
C_{r_1,r_2}(p_1)\left( {\Spab{p_1|\wt r|p_2}^{s+k-1-s_1}\over
\prod_{t=1,t\neq i}^k (\Spab{p_1|\wt q_t|p_2})^{1+z_t}}
+{\Spab{p_2|\wt r|p_1}^{s+k-1-s_1}\over \prod_{t=1,t\neq i}^k
(\Spab{p_2|\wt q_t|p_1})^{1+z_t}}\right). \eea
for which we can apply the general expression in Appendix \eref{Sig-Def}
and \eref{Sig-E}.

\subsection{The Lorentz invariant form}

Putting all together, the coefficient of the  bubble is given by
\bea C[K]&=& \sum_{i=1}^k\sum_{q=0}^{n}{(-1)^{n-q}(K^2)^{2n+1-q}}
\sum_{s=
Max\{n-2q-1,0\}}^{n-q}\sum_{s_1=0}^{s}\sum_{s'_1=0}^{[s_1/2]}
T_0(s,s_1,s_1')T_1(s_1,s_1')\nn &&\left(-{{G(K_i,\wt R,K)}\over K^2
}\right)^{n-q-s}\times \sum_{\stackrel{\sum_{t=1,t\neq i}^k
z_t=s-s_1}{z_t\geq 0}}
\sum_{r_1=0}^{2q+1+s-n}{\sum_{r_2=0}^{n-q-s}}{T_2(z_t)T_4(r_1,r_2)\over
(-4{G(K_i,K)})^{(n-q-s+1+r_1-r_2)/2}}\nn &&\times
\left(2\prod_{t=1,t\neq
i}^k{1\over{G(K_i,K_t,K)}^{1+z_t}}\right)\left(
\sum_{\stackrel{\{h_1,h_2,...,h_k\}\geq 0}{  \sum_{t=1,t\neq i}^k
h_t={\rm even}}}^{\{ 1+z_1,1+z_2,...,1+z_k\}} {
T_3(z_t,h_t)\over {G(K_i,K)}^{n-q-h/2}}\right),~~~\label{Bubb-Coeff} \eea
where
\bea
T_0(s,s_1,s_1')&=&{2^s q!(n-q+k-1)!\over (2q+1+s-n)!s'_1!(s_1-2s'_1)!(n-q+k-1-s_1+s'_1)!(n-q-s)!}\\
T_4(r_1,r_2)\quad&=&{2q+1+s-n \choose r_1}{n-q-s\choose r_2}\left({2\over K^2}G\left(
                               \begin{array}{cc}
                                 K &  K_i \\
                                 K& \wt R \\
                               \end{array}
                             \right)\right)^{r_1}\left({1\over 2K^2}\right)^{r_2}\nn
&&\times(\beta^2(1-u))^{{1\over
2}(n-q-s-1+r_1+r_2)}\alpha_R^{2q+1+s-n-r_1}{(-\alpha_i)}^{n-q-s-r_2}~~~\label{T0T4}
\eea
and $T_1, T_2,T_3 $ are defined in Eq.(\ref{T-N-Def}). In the sum
\eref{Bubb-Coeff}, we need to have $(n-q-s+1+r_1-r_2)$  and $h=\sum_{t=1,t\neq i}^k h_t$
to be even number.

From \eref{Bubb-Coeff} we can see various second-type singularities.
Now we discuss their highest degree. For the pole $G(K_i,K_t,K)$,
there is only one term to contribute. By setting $s=n-q$,
$q=0,s_1=0,r_1=0,1,r_2=0$ we find the highest degree is $(n+1)$. It
is different from the highest degree $(n+2)$ of the same pole in
coefficients of the box and triangle.

For the highest degree of the pole  $K^2$, which is the intrinsic
second-type singularity related to bubble topology, we can find
\bea C[K]\to  {1\over (K^2)^n}\eea
 by
noticing
\bean T_1&\to&{1\over (K^2)^{2s_1}}~~~T_2\to{1\over
(K^2)^{2z_t}}~~\nn T_4&\to&{1\over
(K^2)^{r_1+r_2+(n-q-s-1+r_1+r_2)+2(2q+1+s-n-r_1+n-q-s-r_2)}}\eean

For the pole  $G(K_i,K)=K^2 q_i^2$, its highest degree shows up in
many terms, so we need to rewrite the result to see  clearly.
Using $\delta=-4q_i^2 K^2$ and removing all $q_i^2$ terms in the
numerator we find
\bean C[K]&\to &{(-2)^{n+1}(K^2)^{n}\over \delta^{n+1}}\alpha_i^{n}
\sum_{s=
Max\{n-1,0\}}^{n}\sum_{s_1=0}^{s}\sum_{s'_1=0}^{[s_1/2]}2^{s_1-2s'_1}
T_0(s,s_1,s_1')(-r\cdot q_i)^{k+s}\nn
&&\sum_{\stackrel{\sum_{t=1,t\neq i}^k z_t=s-s_1}{0\leq z_t\leq
s-s_1}} \left(\prod_{t=1,t\neq i}^k\left({(1+z_t)(q_t\cdot
q_i)^{1+2z_t}\over (q_i^2q_t^2-(q_i\cdot
q_t)^2)^{1+z_t}}\right)\right) \eean
From this expression, we can find the highest degree of pole
$G(K_i,K)=K^2 q_i^2$ is $(n+1)$.

\section{Conclusion}

In this paper, to prepare the study of analytic properties of
one-loop amplitudes, we have rewritten the spinor forms of one-loop
coefficients given in \cite{Britto:2007tt} to manifestly
Lorentz-invariant contraction forms of external momenta. Although
the rewriting is a little bit complicated and some skills within the
spinor formalism are needed, the final results of various
coefficients are manageable and have been summarized in the first
subsection of section 4,5,6.

The main results of our calculations are following. Firstly we have
found that although there are two types of singularities by the general
S-matrix analysis, coefficients of each basis contain only {\sl
second-type} singularities while the first-type singularities appears only in the basis.
Secondly, the degree of each second-type
singularities is tightly related to the degree of the inner momentum in
numerators. In other words, when we study the analytic property, not
only the structure of the denominator, but also the structure of
the numerator, play an important role. For the renormalizable theory, there is an up-bound
for the degree of loop momentum $\W\ell$ in the numerator(see the definition of $\W\ell$ in Eq.\eref{Wl-def} and the degree in Eq.\eref{Input}),
thus the possible highest degree
for each second-type singularity is known. For the non-renormalizable theory,
the information of the degree of $\W\ell$ in numerator will also tell us
how bad the contributions from these singularities could be.

Thirdly, for a given basis, its
coefficient contains not only the second-type singularity related to
its topology, but also those related to its
mother topology. Not only that, the degree  also matches up. For
example, the triangle coefficient contains the second-type singularity of
the box topology with the same highest degree $(n+2)$ as the box coefficient,
where $n$ is the difference of the number of loop momentum $\W\ell$ in numerator and the
number of propagators (see Eq.\eref{Input} and below).
This matching  has the physically meaningful cancelation in various
singular limits. To have a clear picture, we have given a table (see
Table \ref{Table-sum}) where for each coefficient, the involved
singularities and their highest degrees have been given.
In this table, the meaning of $n$ can be found in Eq.\eref{Input}  and
$G$, $D$ are defined in Eq.\eref{Gram-def-1} and Eq.\eref{Hex-pen} respectively.
\begin{table}[h]
  \centering \caption{The table of  singularities and their highest
degrees for each coefficient}
  \begin{tabular}{|c|c|c|c|c|c|}\hline
   & $D^{(K_i,K_j,K_t, K_\omega,K)}$ & $G(K_i,K_j,K_t,K)$ &
  $G(K_i,K_j,K)$ &
  $G(K_i,K)$ & $K^2$  \\ \hline {\rm Pentagon} & $1$ & $(n+k)$ & & & \\   \hline
  {\rm Box} & & $(n+3)$ & $(n+2)$ & & \\ \hline {\rm Triangle} & & & $(n+2)$&$(n+1)$ & \\ \hline
  {\rm Bubble} & & &$(n+1)$ & $(n+1)$& $n$ \\ \hline
            \end{tabular}
  ~~\label{Table-sum}
\end{table}

From the table, there is one point we want to mention. The pole
$G(K_i,K_j,K_t,K)$ related to the  pentagon topology shows up only in the
coefficients of the pentagon and box, while the pole $G(K_i,K_j,K)$
related to the  box topology shows up also in the coefficients of the box,
triangle and bubble. We believe the reason is that in the
$(4-2\eps)$-dimension, the pentagon can be expressed as the linear
combination of boxes plus terms in the higher order of $\eps$. Thus
the influence of pentagon can not be propagated to lower topologies.
It is consistent with the well known fact that second-type
singularity depends on the dimension of space-time and structures of
interactions, such as the spins, derivative interactions etc.

We want to emphasize following points for our work. Firstly our paper is to set up
a frame for the theoretical study of analytic property of one-loop
amplitudes. Thus although it will be possible to use these explicit
Lorentz-invariant forms of coefficients in real one-loop
calculations numerically or analytically, we will not do it here and
leave it as a future project. To use our result in real
calculations, we need to discuss the efficiency or stability of
calculations as carefully discussed, for example, in paper
\cite{Denner:2005nn}.

One possible application of our results is following.  Since our
results are complete, i.e., there are $(\mu^2)^n$-terms
corresponding to rational part as mentioned in section 2, we could use
our result to calculate the rational terms and compare with results from
 the recursion relation  given in \cite{Bern:2005hs,Bern:2005ji,Bern:2005cq}.
We believe this calculation will help to us to clarify some points in the
recursion relation.

Secondly, our current focus is coefficients of various basis, not
the whole structure of one-loop amplitudes. For the latter, one need
to address cancelations of various singularities under various
limits (such as soft and collinear limits) by combining
contributions from various basis. The cancelation of these
singularities in a complete loop amplitudes is very intricate. It reflects
many important information of the theory under consideration, such as how
good (or bad) the divergence will be and if the theory has some unexpected
hidden symmetries. It
is definitely an important issue and relates to the application in real
processes as mentioned in previous paragraph. Although our results
in this paper provide a starting point for these discussions, its
explicit demonstration will be very complicated and deserves to be
an independent work.

Thirdly, for coefficients of boxes, triangles and bubbles, there are many terms,
thus there are many different ways to write down the sum. However, we want
to emphasize although there are many different ways  to group numerators,
the denominators are the same. In other words, the appearance of various
second-type singularities  is common for all different expressions, especially
the highest degrees  of second-type singularities are the same.
The choice we made here is because we believe  this choice gives the
best presentation of singularity structure.

Fourthly, as we have mentioned again and again,
the coefficients of various bases contain only
second-type singularities as classified in \cite{S-matrix, Second}.
One exception is the singularities for the pentagon as given in
$D^{K_i,K_j,K_t,K_\omega,K}$ ( Eq.(\ref{Pen-coeff})). Although it
contains the mass, we do not think it belongs to the first-type
singularity. This exception is related to the special position of pentagon
as a basis in the PV-reduction as we have discussed many times in the paper.
What it means or if it is a really a singularity
deserves further study.

Fifthly, although in this paper, we have identified all
singularities. It is still have a lot of work to do to understand
their properties. For example, we need to know  if they are true
singularities for one-loop amplitudes when we sum all together. If
they are,  where are their locations: on the physical sheet or
unphysical sheet. Here we want to distinguish one thing. One of our
motivation of current calculations is to find a recursive way to
calculate coefficients of various basis. Thus all singularities we
found in the paper do contribute coefficients no matter whether they
are physical or not or which sheet they locate at. The situation is
different from BCFW on-shell recursion relation for tree-level
amplitudes where we calculate the complete (global) tree amplitudes
thus spurious poles do not give contributions.

Sixthly, another possible application (which is one
of our main motivations) of our results should be mentioned.
With Lorentz-invariant forms of
coefficients, we can study their factorization property under the
various deformation (such as the BCFW-deformation or Risager's deformation \cite{Risager:2005vk}).
Based on information under deformation,  we can try if it is possible to
establish some sort of recursion relation.

Finally we would like to give a general remark about the unitarity cut method\footnote{
We would like to thank the referee for suggestions and remarks.}.
This method has several advantages comparing with other methods. For example,
the input is  the product of on-shell tree-level amplitudes, so the expressions can be
very compact. Also we do not need to calculate coefficients of spurious bases
opposite to the powerful OPP-method \cite{Ossola:2006us}. However, the unitarity cut method
has some limitations. For example, it cannot be applied to the calculation of on-shell fermion
self-energy, where one denominator is massless, and the
second denominator has a mass equal to the mass of the external leg (see \cite{Britto:2011cr}).
Another  problem for the unitarity cut method is that it can not detect
tadpole coefficients directly. Tadpole coefficients
can exist as long as the inner propagator is massive.  To deal with tadpole coefficients, several methods have
been suggested, for example, the single cut method \cite{Britto:2011cr, Britto:2009wz}
and using the universal UV and IR behavior as did in \cite{Bern:1995db,Badger:2008za}.
Since we have analytic expressions for all other coefficients, it will be very interesting
to see if one can use the universal UV and IR behavior to find the general analytic expression.

\section*{Acknowledgements}

This project  is supported, in part, by fund from Qiu-Shi and
Chinese NSF funding under contract No.11031005, No.11135006, No.
11125523.

\appendix

\section{Sum in spinor form to Lorentz form}
In this appendix, we will present a formula which  is very important
to transform the sum in the spinor form to the Lorentz-invariant form.
The typical sum we meet again and again is the following
\bea
 \Sigma_N\equiv{\Spab{P_1|T|P_2}^N\over \prod_{t=1}^N\Spab{P_1|Q_t|P_2}}
 +{\Spab{P_2|T|P_1}^N\over \prod_{t=1}^N\Spab{P_2|Q_t|P_1}}~~~\label{Sig-Def}
\eea
and
{\small\bea \Sigma_{N-1}[Q_m]\equiv{\Spab{P_1|T|P_2}^N\over
\prod_{t=1,t\neq m}^N \Spab{P_1|Q_t|P_2}}+{\Spab{P_2|T|P_1}^N\over
\prod_{t=1,t\neq m}^N\Spab{P_2|Q_t|P_1}},
~\Sigma_{1}(Q_m)\equiv{\Spab{P_1|T|P_2}\over
\Spab{P_1|Q_m|P_2}}+{\Spab{P_2|T|P_1}\over
\Spab{P_2|Q_m|P_1}}.~~~\label{Sig-Def-1} \eea}
where $P_1$ and $P_2$ are two massless momenta constructed from $Q_i$
and $Q_j$ (see Eq.\eref{P1P2} for explicit construction). Furthermore we suppose $i$ and $j$ are not in the set
$\{1,\ldots,N\}$ of (\ref{Sig-Def}). Our derivation of the Lorentz-invariant
form will use the inductive method.

\subsection{Recursion relation}
For a given pair $(n,m)$ simple calculations from (\ref{Sig-Def-1})
give
\bea
\Sigma_1(Q_n)\Sigma_{N-1}[Q_n]&=&\Sigma_N+{\Spab{P_1|T|P_2}\Spab{P_2|T|P_1}\Spab{P_2|T|P_1}^{N-2}\over
\Spab{P_1|Q_n|P_2}\Spab{P_2|Q_m|P_1}\prod_{t=1,t\neq
n}^{N-1}\Spab{P_2|Q_t|P_1}}\nn &&+
{\Spab{P_2|T|P_1}\Spab{P_1|T|P_2}\Spab{P_1|T|P_2}^{N-2}\over
\Spab{P_2|Q_n|P_1}\Spab{P_1|Q_m|P_2}\prod_{t=1,t\neq n}^{N-1}\Spab{P_1|Q_t|P_2}}\label{S-1}\\
\Sigma_1(Q_m)\Sigma_{N-1}[Q_m]&=&\Sigma_N+{\Spab{P_1|T|P_2}\Spab{P_2|T|P_1}\Spab{P_2|T|P_1}^{N-2}\over
\Spab{P_1|Q_m|P_2}\Spab{P_2|Q_n|P_1}\prod_{t=1,t\neq
m}^{N-1}\Spab{P_2|Q_t|P_1}}\nn &&+
{\Spab{P_2|T|P_1}\Spab{P_1|T|P_2}\Spab{P_1|T|P_2}^{N-2}\over
\Spab{P_2|Q_m|P_1}\Spab{P_1|Q_n|P_2}\prod_{t=1,t\neq
m}^{N-1}\Spab{P_1|Q_t|P_2}}.\label{S-2} \eea
The sum of Eq. (\ref{S-1}) and  Eq. (\ref{S-2}) yields
\bean \Sigma_1(Q_n)\Sigma_{N-1}[Q_n]+\Sigma_1(Q_m)\Sigma_{N-1}[Q_m]
&=&2\Sigma_N+\left({\Spab{P_1|T|P_2}\Spab{P_2|T|P_1}\over
\Spab{P_1|Q_n|P_2}\Spab{P_2|Q_m|P_1}}+{\Spab{P_1|T|P_2}
\Spab{P_2|T|P_1}\over \Spab{P_1|Q_m|P_2}\Spab{P_2|Q_n|P_1}}\right)\\
&\times&\Sigma_{N-2}[Q_n,Q_m]. \eean
Using the spinor formulism (remembering $P_1, P_2$ are two massless
momenta constructed from $Q_i, Q_j$), after some trivial
manipulations we can get
\bea
\Spab{P_1|Q_n|P_2}\Spab{P_2|Q_m|P_1}+\Spab{P_1|Q_m|P_2}
\Spab{P_2|Q_n|P_1}&=&-{8\over Q_i^2}(Q_n|Q_m)\\
\Spab{P_1|T|P_2}\Spab{P_2|T|P_1}&=&-{4\over Q_i^2}(T|T)
\eea
where we have defined
\bea
(Q_n|Q_m)\equiv \det\left(
 \begin{array}{ccc}
  Q_i^2 & Q_i\cdot Q_j & Q_i\cdot Q_n \\
    Q_i\cdot Q_j & Q_j^2 & Q_j\cdot Q_n \\
      Q_i\cdot Q_m & Q_j\cdot Q_m & Q_n\cdot Q_m\\
      \end{array} \right)~.\label{Q-Q-def}
\eea
So we get the following  relation
 \bea
\Sigma_1(Q_n)\Sigma_{N-1}[Q_n]+\Sigma_1(Q_m)\Sigma_{N-1}[Q_m]=2\Sigma_N+2{(T|T)(Q_n|Q_m)\over
(Q_n|Q_n)(Q_m|Q_m)}\Sigma_{N-2}[Q_n,Q_m]~~~\label{Sig-1} \eea

Summing over all pairs $(n,m)$ of (\ref{Sig-1}) we get
\bea (N-1)\sum_{t=1}^{N}\Sigma_1(Q_t)\Sigma_{N-1}[Q_t]
&=&N(N-1)\Sigma_N+2\sum_{1\leq t<k \leq N}{(T|T)(Q_k|Q_t)\over
(Q_k|Q_k)(Q_t|Q_t)}\Sigma_{N-2}[Q_k,Q_t] \eea
or
\bea \Sigma_N={1\over
N}\left(\sum_{t=1}^{N}\Sigma_1(Q_t)\Sigma_{N-1}[Q_t] -{2\over
N-1}\sum_{1\leq t<k \leq N}{(T|T)(Q_k|Q_t)\over
(Q_k|Q_k)(Q_t|Q_t)}\Sigma_{N-2}[Q_k,Q_t]\right)\label{R-R} \eea
%
\subsection{Proof by inductive method}

With some calculations, we find the explicit expression of
$\Sigma_N$ to be
{\small\bea \Sigma_N &=&{1\over
\prod_{k=1}^N(Q_k|Q_k)}\left(2^N\prod_{j=1}^N(T|Q_j)\right.\nn &&
\left. +\sum_{m=1}^{[N/2]}{(-1)^m
2^{N-m}m!(N-2m)!A_{N,m}(T|T)^m\over N!}\sum_{m~{\rm
pairs}}\prod_{p=1}^{m}(Q_{p_1}|Q_{p_2}) \prod_{q\in \{N\}-\{m~{\rm
pairs}\}}^{N}(Q_{q}|T)\right).~~\label{Sig-E} \eea}
where the notation $[N/2]$ means to take the maximum integer equal
to or less than $N/2$, and
\bea  A_{n,m}=\left\{ \begin{array}{ll} A_{n-1,m}+A_{n-2,m-1},~~ &
2m<n \\  2, & 2m=n \\ 0, & 2m>n \end{array} \right.
~~~\label{A-def}\eea
 The second sum at the second line of (\ref{Sig-E}) is over
all different choices of $m$ pairs in the set $\{1,2,...,N\}$ and
each pair contributes a factor $(Q_{p_1}~Q_{p_2})$. After $m$ pairs
having been chosen, each remaining element  will contribute a factor
$(Q_{q}~T)$. First few examples $N=1,2,3$ can be calculated directly
as
\bean \Sigma_1&=&{\Spab{P_1|T|P_2}\over
\Spab{P_1|Q_1|P_2}}+{\Spab{P_2|T|P_1} \over
\Spab{P_2|Q_1|P_1}}=2{(T|Q_1)\over (Q_1|Q_1)},\\
\Sigma_2&=&2^2{(T|Q_1)(T|Q_2)\over (Q_1|Q_1)(Q_2|Q_2)}-2{(T|T)(Q_2|Q_1)\over (Q_1|Q_1)(Q_2|Q_2)},\\
\Sigma_3&=&{1\over
(Q_1|Q_1)(Q_2|Q_2)(Q_3|Q_3)}(2^3(T|Q_1)(T|Q_2)(T|Q_3)-2(T|T)(Q_2|Q_1)(T|Q_3)\nn
&&-2(T|T)(Q_3|Q_1)(T|Q_2)-2(T|T)(Q_3|Q_2)(T|Q_1)),\eean
which are the same as given by (\ref{Sig-E}).

We will prove the formula (\ref{Sig-E}) by showing that it satisfies
the relation Eq. (\ref{R-R}) by inductive method. We check this term by
term. For the first term of Eq.({\ref{Sig-E}}), it satisfies
the relation Eq. (\ref{R-R}) obviously since only the first term of Eq.
(\ref{R-R}) contributes. For the second part of the formula
(\ref{Sig-E}) with given $m$ pairs in the set  $\{1,2,...,N\}$, both
terms of Eq. (\ref{R-R}) will contribute. To simplify our
discussion, we use the set ${\cal M}=\{ m~{\rm pairs}\}$ and the set
${\cal Q}= \{ N\}- {\cal M}$. The contribution from the first term
of Eq. (\ref{R-R}) is given by
\bean T_1& = & {(-1)^m 2^{N-1-m}m!(N-1-2m)!A_{N-1,m}(T|T)^m\over N
\prod_{k=1}^N(Q_k|Q_k)(N-1)!}\prod_{p\in {\cal M}}(Q_{p_1}|Q_{p_2})
\sum_{q\in {\cal Q}} (T|Q_q) \prod_{\W q\in
{\cal Q}-q}(Q_{\W q}|T)\\
& = & {(-1)^m 2^{N-1-m}m!(N-1-2m)!A_{N-1,m}(T|T)^m\over N
\prod_{k=1}^N(Q_k|Q_k)(N-1)!}\prod_{p\in {\cal M}}(Q_{p_1}|Q_{p_2})
(N-2m) \prod_{ q\in {\cal Q}}(Q_{ q}|T).\eean
The sum in the first line of $T_1$ comes from choosing which $q\in
{\cal Q}$ belongs to the $\Sigma_1$ part. The contribution from the
second term of Eq. (\ref{R-R}) is given by
\bean T_2& = &-{(-1)^{m-1}
2^{N-m-1}(m-1)!(N-2m)!A_{N-2,m-1}(T|T)^m\over N (N-1)
\prod_{k=1}^N(Q_k|Q_k) (N-2)!}\prod_{ q\in {\cal Q}}(Q_{
q}|T)\sum_{p\in {\cal M}} (Q_{p_1}|Q_{p_2})\prod_{\W p\in
{\cal M}-p}(Q_{\W p_1}|Q_{\W p_2}) \\
& = & -{(-1)^{m-1} 2^{N-m-1}(m-1)!(N-2m)!A_{N-2,m-1}(T|T)^m\over
N(N-1) \prod_{k=1}^N(Q_k|Q_k) (N-2)!}\prod_{ q\in {\cal Q}}(Q_{
q}|T)m\prod_{\W p\in {\cal M}}(Q_{\W p_1}|Q_{\W p_2}).\eean
The sum in the first line of $T_2$ comes from choosing which pair
$p\in {\cal M}$ does not belong to the $\Sigma_{N-2}$ part. Summing
$T_1$ and $T_2$, with a little algebra, we can see that it
reproduces the corresponding terms of formula (\ref{Sig-E}).

A special case of the above proof is that when $N=2m$, only the second
term of Eq. (\ref{R-R}) contributes. It is easy to see that we do
have $A_{2m,m}=A_{2m-2,m-1}=2$ as given by (\ref{A-def}).


\end{document}